\documentclass[%
 reprint,
 superscriptaddress,
 amsmath,amssymb,
 aps,
 prl,
]{revtex4-2}

\usepackage{physics}
\usepackage{graphicx}
\usepackage{dcolumn}
\usepackage{bm}

\usepackage{xcolor}

\begin{document}

\title{On the origin of bulk-related anisotropies in surface optical spectra}

\author{Max Großmann}
\email{max.grossmann@tu-ilmenau.de}
\affiliation{Theoretical Physics I, Institute of Physics, Technische Universit\"at Ilmenau, 98693 Ilmenau, Germany}
\affiliation{Center of Micro- and Nanotechnologies, Technische Universit\"at Ilmenau, 98693 Ilmenau, Germany}

\author{Kai Daniel Hanke}
\affiliation{Center of Micro- and Nanotechnologies, Technische Universit\"at Ilmenau, 98693 Ilmenau, Germany}
\affiliation{Fundamentals of Energy Materials, Institute of Physics, Technische Universit\"at Ilmenau, 98693 Ilmenau, Germany}

\author{Chris Yannic Bohlemann}
\affiliation{Center of Micro- and Nanotechnologies, Technische Universit\"at Ilmenau, 98693 Ilmenau, Germany}
\affiliation{Fundamentals of Energy Materials, Institute of Physics, Technische Universit\"at Ilmenau, 98693 Ilmenau, Germany}

\author{Agnieszka Paszuk}
\affiliation{Center of Micro- and Nanotechnologies, Technische Universit\"at Ilmenau, 98693 Ilmenau, Germany}
\affiliation{Fundamentals of Energy Materials, Institute of Physics, Technische Universit\"at Ilmenau, 98693 Ilmenau, Germany}
\affiliation{BMBF Junior Research Group PARASOL, Institute of Physics, Technische Universit\"at Ilmenau, 98693 Ilmenau, Germany}

\author{Thomas Hannappel}
\affiliation{Center of Micro- and Nanotechnologies, Technische Universit\"at Ilmenau, 98693 Ilmenau, Germany}
\affiliation{Fundamentals of Energy Materials, Institute of Physics, Technische Universit\"at Ilmenau, 98693 Ilmenau, Germany}

\author{Wolf Gero Schmidt}
\affiliation{Lehrstuhl f\"ur Theoretische Materialphysik, Universit\"at Paderborn, 33095 Paderborn, Germany}

\author{Erich Runge}
\affiliation{Theoretical Physics I, Institute of Physics, Technische Universit\"at Ilmenau, 98693 Ilmenau, Germany}
\affiliation{Center of Micro- and Nanotechnologies, Technische Universit\"at Ilmenau, 98693 Ilmenau, Germany}

\date{\today}

\begin{abstract}
Reflection anisotropy spectroscopy (RAS) is a powerful method for probing the optical properties of surfaces, used routinely in research and industrial applications, yet the origin of 'bulk-related' features that appear in the spectra of various surfaces has been debated for nearly 40 years.
It is often argued that these features are related to surface-induced bulk anisotropy (SIBA) because they coincide with critical energies of the bulk dielectric function.
In general, any quantitative RAS theory must include excitonic effects as they significantly influence the spectra and are believed to be the key to determining the origin of SIBA features.
Here, we introduce a layer-resolved exciton localization (LREL) measure within the framework of many-body perturbation theory, which enables a quantitative analysis of the origins of 'bulk-related' RAS features.
Applying LREL to arsenic-modified silicon reconstructions reveals that, depending on the surface reconstruction, the 'apparent' SIBA features arise primarily from states localized at the surface, with only a small contribution from the underlying layers.
Our findings, further supported by the fact that the calculated spectra agree well with low-temperature RAS measurements, challenge the conventional explanation of 'bulk-related' RAS features.
They indicate that in many instances bulk-enhanced surface anisotropies (BESA)---the opposite of SIBA---contribute to, or are even responsible for, ’bulk-related’ RAS features.
Therefore, we suggest that previously studied semiconductor surfaces, which exhibit 'bulk-related' features in their spectra, should be reanalyzed using the presented method.
\end{abstract}

\maketitle

\section{Introduction}

Reflection anisotropy (RA) spectroscopy (RAS) has been extensively studied and discussed for a variety of semiconductor surfaces and heterostructures over the last decades due to its importance in surface characterization and technological applications \cite{Aspnes1985, Mochan1985, Kipp1996, Uwai1997, Rohlfing1999, Sheridan2000, Hahn2001, Chandola2009, Kumar2017, Kress1997, Schmidt2001, Hogan2003, Fuchs2005, Chandola2009, Uwai1997, Zorn1999, Hahn2003, Fazi2013, Hogan2018, Palummo1999, WGS-layer, Schmidt2003, Landmann2015, Alvarado2023, Palummo2009, HINGERL2001769, DelSole1999, Arciprete2003, Hogan2004, Hingerl2000, Schmidt2000, WGS-E-Feld, AcostaOrtiz1989}. 
However, the interpretation of RAS experiments has generally been far from straightforward, as RA spectra give only indirect insights into surface structure and chemistry. 
Furthermore, interpreting spectra based on simplified models, such as bond polarizability is difficult and may lead to wrong results \cite{Schmidt2004}.
Arguably, the full potential of RAS can therefore only be realized through a strong collaboration between experimental and theoretical work, especially in understanding the complex origins of the spectral features.

Today, RAS signatures due to transitions between surface states are well understood and have been enormously useful in surface exploration for decades, as reviewed in detail by Weightman~\textit{et al.}~\cite{Weightman2005}.
However, a topic that has been observed for many surfaces and has been heavily debated for the last 40 years (e.g., Refs.~\cite{Aspnes1985, Aspnes1998, Mantese1999, Mantese1999b, Rossow2000}) is the origin of 'bulk-related' features in RA spectra, i.e., features that coincide with critical points of the imaginary part of the bulk dielectric function, also referred to as surface-induced bulk anisotropy (SIBA).
Irrespective of the investigated surface or specific reconstruction, e.g, Si(100)-$c(4 \times 2)$, Si(100)-$(2 \times 1)$:H and Si(100)-$(1 \times 1)$:H \cite{Palummo2009}, Si(111):H and Si(110):H \cite{HINGERL2001769}, GaAs(110) and GaAs(100)-$\beta2(2 \times 4)$ \cite{DelSole1999}, GaAs(100)-$c(4 \times 4)$ \cite{Arciprete2003}, As-rich GaAs(100)-$c(2 \times 8)$ \cite{Hogan2004}, ZnTe(100) \cite{Hingerl2000} and InP(100)-$c(2 \times 4)$ \cite{Schmidt2000} the occurrence of SIBA features has been noted. 
Their origin is often argued to be related to "bulk-like electronic states perturbed by the surface" \cite{Schmidt2000}. 
Specifically, it has been proposed that they may be related to strain induced by surface reconstructions, steps or dimerization via the piezo-optic effect \cite{HINGERL2001769} or to surface electric fields via the linear electro-optic effect \cite{AcostaOrtiz1989, WGS-E-Feld}. 
Aspnes~\textit{et al.}~\cite{Aspnes1998} suggested that they are caused by photon-induced localization of bulk electronic states. 
In fact, some authors \cite{DelSole1999} have gone so far as to state: "after realizing that many RA line shapes are similar to the imaginary part of the bulk dielectric function, or to its energy derivative, it is a growing belief that surface optical spectra are mostly determined by bulk effects, and therefore, not very useful as a tool of surface characterization." 
While this sentiment has softened over the years, a clear explanation of SIBA features remains elusive.

From a theoretical point of view, it has long been recognized that excitonic effects, which are not captured by the independent particle approximation (IPA), contribute significantly to the RAS of semiconductor surfaces \cite{Rohlfing1999, Hahn2001, Schmidt2003} and may be the key to explaining bulk-related features \cite{DelSole1999}.
Nevertheless, only a limited number of studies \cite{Rohlfing1999, Hahn2001, Schmidt2003, Palummo2009, Landmann2015, Alvarado2023} have addressed these excitonic contributions by solving the Bethe-Salpeter equation (BSE) within the framework of many-body pertubation theory (MBPT) \cite{Onida2002}.
A detailed analysis of the excitonic transitions using BSE eigenvectors \cite{Albrecht1998, Fuchs2008} analogous to recent work on excitons near the bandgap of bulk or two-dimensional materials \cite{Gorelov2022, Acharya2023, Varrassi2024} remains to be fully explored for surfaces.
The main challenge has been the computational cost associated with the exact diagonalization of the large, nonsparse BSE-Hamiltonian matrix \cite{Hahn2001, Schmidt2003, Palummo2004}.
As a result, previous investigations have either restricted the transition space, facilitated by a favorable surface band structure \cite{Rohlfing1999, Landmann2015}, or employed iterative diagonalization algorithms \cite{Hahn2001, Schmidt2003, Palummo2009}, which yield the RAS without elucidating the underlying origins of the excitonic features.
Even when the BSE was exactly diagonalized \cite{Alvarado2023}, a characterization of the excitonic features of RA spectra was not performed.

A detailed interpretation of excitonic features, especially for the aforementioned SIBA features, through a precise methodology would be particularly important for experiments monitoring RAS changes during sample growth or exposure experiments \cite{May2013, May2014, May2017expo, Ostheimer2024}.
To address this issue, we introduce the layer-resolved exciton localization (LREL) measure as a conceptual and quantitative framework for analyzing excitonic contributions to RA spectra.
LREL combines the BSE eigenvectors with the state- and layer-resolved projected wavefunction characters, allowing us to specify from which atomic layer the excitonic contributions to the RAS originate.

While previous studies have been limited by the daunting computational demands of diagonalizing the BSE for surfaces, advances in high performance computing resources are finally making it possible to clarify the origin of the SIBA features:
They allow us to accurately diagonalize the BSE and rigorously post-process the results using our LREL measure for large unit cells, i.e., the surfaces studied here.
As a result, a detailed analysis of excitonic effects in semiconductor surfaces becomes possible, shedding light on the elusive origin of SIBA features.
Thus, our study not only addresses a long-standing problem in the field, but also provides a recipe for the thorough analysis of surface optical spectra.

\begin{figure*}[ht]
    \centering
    \includegraphics{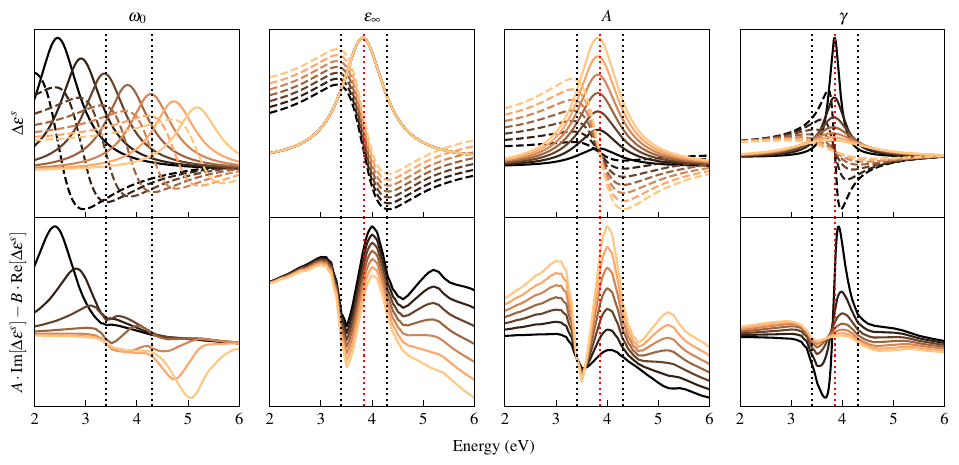}
    \caption{
    Illustration of the effect of varying the surface dielectric anisotropy on the RAS, using silicon as a fixed bulk material with $\varepsilon^b$ taken from Aspnes~\textit{et al.}~\cite{Aspnes1983}.
    The top row shows the surface dielectric anisotropy models (see text), with each column showing variations in a specific model parameter, as indicated by the column titles.
    The imaginary and real parts of the surface dielectric anisotropy are represented by solid and dashed lines, respectively.
    The bottom row shows how the RAS, i.e., Eq.~(\ref{eq:ras_split}) without the prefactor $4\pi \omega d c^{-1}$, changes as the surface dielectric anisotropy is varied.
    The line colors change linearly from black to orange with increasing values of the parameter being varied.
    The critical point energies of bulk silicon \cite{Lautenschlager1987} are marked by vertical dotted black lines in all panels. 
    In columns 2-4, a dashed red line marks the value of $\omega_0$ used for the surface dielectric anisotropy.
    }
    \label{fig1}
\end{figure*}

To test our method on a real system with a more challenging surface band structure than, for example, the Si(111)-($2\times 1$) reconstruction~\cite{Rohlfing1999}, we focus on arsenic-modified Si(100) surfaces, which exhibit strong SIBA features. 
These surfaces have recently gained attention as a promising substrate for low-defect III-V-on-Si heteroepitaxy \cite{Supplie2017,Supplie2018,Paszuk2018a,Paszuk2018b,Nandy2021}. 
They have also been shown to be crucial for the development of low-cost, high-efficiency photoelectrochemical devices \cite{May2017, Alqahtani2019, Feifel2021, Hannappel2024}.

We focus on two particular Si(100) reconstructions that form during arsenic exposure under different growth conditions. 
The first is the Si(100)-$(2\times 1)$:As reconstruction with two symmetric As dimers, previously studied by Kipp~\textit{et al.}~\cite{Kipp1996}, which is formed during an ultra-high vacuum (UHV) preparation using As$_4$ precursors \cite{Uhrberg1986, Kipp1996}.
The second is the Si(100)-$(2\times 1)$:As-Si-H reconstruction, which we have recently demonstrated to occur in chemical vapor deposition (CVD) environments, where large amounts of background hydrogen and hydrogen from tertiarybutylarsine (TBA) precursors lead to the formation of a reconstruction with mixed buckled As-Si-H dimers \cite{Bohlemann2024}.
As experimental reference, we use data measured by Kipp~\textit{et al.}~\cite{Kipp1996} at room temperature (RT) for the Si(100)-$(2\times 1)$:As reconstruction and performed our own low-temperature (LT) RAS measurements for the Si(100)-$(2\times 1)$:As-Si-H reconstruction, described in the Methods section.
For the sake of brevity, we will omit $(2\times 1)$ from the notation of the surface reconstructions from now on, since all surfaces considered here share this unit cell.

Before introducing our analysis method and presenting results, we first outline and discuss the equations commonly used in the analysis of RA spectra.
In general, RAS measures the difference between complex Fresnel reflection amplitudes $r$ along two orthogonal directions $x$ and $y$ in the surface plane of a sample.
\begin{equation}\label{eq:ras}
    \frac{\Delta r}{r} = \frac{2(r_x-r_y)}{r_x + r_y}
\end{equation} 
Experimentally, the real part of the RAS is usually measured, i.e., $\mathrm{Re}(\Delta r/r)$,
and often used directly for comparison with \textit{ab initio} calculations for $\Delta R/R$, which refer to the reflectivity $R = \vert r^2\vert$. 
However, this direct comparison can lead to overestimated anisotropies because $2\,\mathrm{Re}[\Delta r/r] \approx \Delta R/(R)$ \cite{Schmidt2005}. 
Taking into account this factor of two, we can calculate the RAS for normally incident light using the equation established in the works of Del~Sole~\textit{et al.}~\cite{DelSole1981, DelSole1984} and Manghi~\textit{et al.}~\cite{Manghi1990}:
\begin{equation}
    \label{eq:ras_ai}
    \mathrm{Re}\left[\frac{\Delta r}{r}\right] = \frac{8\pi\omega}{c}\,\mathrm{Im}\left[\frac{\Delta\alpha^\mathrm{hs}(\omega)}{\varepsilon^b(\omega)-1}\right]
\end{equation}
Here, $\varepsilon^b(\omega)$ is the bulk dielectric function of the substrate, and $\Delta \alpha^\mathrm{hs}(\omega)$ is the half-slab polarizability of the (reconstructed) surface. 
From an \textit{ab initio} calculation of a symmetric surface slab approximating a semi-infinite crystal, $\Delta \alpha^\mathrm{hs}(\omega)$ is obtained from the difference of the diagonal components of the surface dielectric tensor $\varepsilon_{ii}^s(\omega)$, i.e., $\Delta \alpha^\mathrm{hs}(\omega) = d\,\Delta\varepsilon^s/2 = d\,[\varepsilon_{xx}^s(\omega) - \varepsilon_{yy}^s(\omega)]/2$, where $d$ is half the thickness of the surface slab \cite{Manghi1990} and mirror symmetry $z \leftrightarrow -z$ or inversion symmetry $\mathbf{r}\leftrightarrow-\mathbf{r} $ is assumed \cite{Hogan2003}.
From here on, we will refer to $\Delta\varepsilon^s$ as the surface dielectric anisotropy. 

At this point, it is beneficial to examine Eq.~(\ref{eq:ras_ai}) in more detail to understand the effect of the bulk dielectric function on RA spectra. 
First, we decompose Eq.~(\ref{eq:ras_ai}) into the real and imaginary parts of bulk and surface terms \cite{Selci1987}
\begin{equation}
    \label{eq:ras_split}
    \mathrm{Re}\left[\frac{\Delta r}{r}\right] = \frac{4\pi \omega d}{c} \big(A\cdot\mathrm{Im}[\Delta\varepsilon^s] - B\cdot\mathrm{Re}[\Delta \varepsilon^s]\big)
\end{equation}
where
\begin{align}
    A &= \frac{\mathrm{Re}[\varepsilon^b] - 1}{(\mathrm{Re}[\varepsilon^b]-1)^2 + (\mathrm{Im}[\varepsilon^b])^2}\nonumber\\
    B &= \frac{\mathrm{Im}[\varepsilon^b]}{(\mathrm{Re}[\varepsilon^b]-1)^2 + (\mathrm{Im}[\varepsilon^b])^2}.\nonumber 
\end{align}
Obviously, the surface dielectric anisotropy $\Delta \varepsilon^s$ can be strongly modulated by the bulk contributions $A$ and $B$ when the real part of the bulk dielectric is close to one while the imaginary part is small. 
For example, in bulk silver (see Fig.~4 in Ref.~\cite{Hogan2018}), this phenomenon occurs around $\omega=3.8$~eV, which coincides with the point where the real part of the dielectric function approaches one, and still before the onset of the $d^{10}s^1 \rightarrow d^9s^2$ absorption peak is observed in the imaginary part. 
At this energy, $\mathrm{Im}[\varepsilon^b]$ remains relatively small.

To further explore the influence of the bulk dielectric function on the RAS, we introduce a simple model system with silicon as the bulk (substrate) material.
It illustrates explicitly that spectral features of a completely \textit{isotropic} bulk material can show up as frequency-dependent reflection anisotropy of the surface!

In our model calculations, we systematically vary the surface dielectric anisotropy of a fictitious surface layer on top of isotropic bulk silicon. 
The latter is modeled via the dielectric function $\varepsilon^b$ and $A$, $B$ measured by Aspnes~\textit{et al.}~\cite{Aspnes1983}.
We note in passing that the structures of $A$ and $B$ are similar to those of other common semiconductors such as Ge, GaAs, and GaP (see Ref.~\cite{Selci1987}). 
The surface dielectric anisotropy is modeled as a single Lorentz oscillator 
$\varepsilon(\omega) = \varepsilon_\infty + A[(\omega_0^2 - \omega^2) - \mathrm{i}\gamma\omega]^{-1}$, 
with $\varepsilon_{yy}^s(\omega) = 0.9\,\varepsilon_{xx}^s(\omega)$, i.e., $\Delta \varepsilon = 0.1\,\varepsilon_{xx}^s$, and we explore the influence of the four model parameters ($\omega_0$, $\varepsilon_\infty$, $A$, $\gamma$) on the RAS.
The numerical values of the model parameters and a visualization of $A$ and $B$ derived from the measured bulk dielectric function are given in the Supplementary Information, i.e., Supplementary~Note~1.
Our model can be interpreted as accounting for changes in the electronic structure at the surface of a semiconductor substrate, such as those resulting from changes in surface reconstruction due to varying growth conditions or the addition of a monolayer with anisotropic optical properties.

The model surface dielectric anisotropy and the resulting RAS, i.e., Eq.~(\ref{eq:ras_split}) without the prefactor $4\pi \omega d c^{-1}$, are shown in Fig.~\ref{fig1}.
Looking at the first column, where we vary $\omega_0$ while keeping the other parameter fixed, we see that the sign of the RAS changes depending on whether the surface dielectric anisotropy is below or above the critical points of the bulk dielectric function (indicated by the dotted black lines). 
When the peak of the surface dielectric anisotropy is at or between the critical points of the bulk dielectric function ($\omega_0 \in [3.4,\, 4.3]$~eV), a strong modulation of the RAS is seen, which could easily and erroneously be attributed to SIBA rather than to what could be called 'bulk-enhanced surface anisotropy' (BESA).
In columns 2-4 in Fig.~\ref{fig1}, the other model parameters are varied, while keeping $\omega_0 = 3.85$~eV fixed (dotted red lines).
Generally, negative peaks are observed near the critical points of the bulk dielectric function, and the strong (positive) peak attributed to the surface dielectric anisotropy is slightly blue-shifted. 
Varying the other model parameters while keeping $\omega_0$ fixed shows that the position of the surface dielectric anisotropy feature has the most significant influence on the RAS, while other parameters primarily affect the amplitude.
In summary, the model results demonstrate that the surface dielectric anisotropy may be strongly modulated by the bulk dielectric function, depending on the relative position of bulk and surface spectral features. 
This interaction complicates the interpretation of RA spectra, especially when investigating SIBA features. 
Therefore, a more comprehensive analysis of RA spectra incorporating the surface dielectric anisotropy, derived from \textit{ab initio} calculations is called for.

In an \textit{ab initio} context and taking into account excitonic effects (electron-hole interactions), the surface dielectric anisotropy is calculated using the surface dielectric tensor, which is obtained here by solving the BSE in the Tamm-Dancoff approximation \cite{Bechstedt2014} as a generalized eigenvalue problem \cite{Fuchs2008}.
The diagonal components of the surface dielectric tensor are then calculated as follows:
\begin{align}\label{eq:bse_eps}
    \varepsilon_{ii}^s(\omega)= 1+\frac{8 \pi e^2 \hbar^2}{m_0^2}\sum_\Lambda O^\Lambda_{ii}  \sum_{\beta = \pm 1}\frac{1}{E^\Lambda-\hbar\beta(\omega+i\gamma)}
\end{align}
where $E^\Lambda$ are the exciton excitation energies (BSE eigenvalues) and $O^\Lambda_{ii}$ is the oscillator strength along axis $i$.
Here $\gamma$ is a phenomenological broadening parameter that accounts for the finite lifetime and damping of excitonic states.
In practice, this is just a small real number, see Methods.
The oscillator strength $O^\Lambda_{ii}$ is defined as:
\begin{equation}\label{eq:osci_str}
    O^\Lambda_{ii} = \frac{1}{\Omega} \left\vert\sum_{vc\mathbf{k}}F_{vc\mathbf{k}}^{i} A^\Lambda_{vc\mathbf{k}}\right\vert^2  
\end{equation}
where $A^\Lambda_{vc\mathbf{k}}$ are the normalized excitonic wave function coefficients (BSE eigenvectors) obtained from a diagonalization of the BSE Hamiltonian \cite{Fuchs2008} with $\sum_{vc\mathbf{k}}\vert A^\Lambda_{vc\mathbf{k}} \vert^2 = 1$, and $\Omega$ is the unit cell volume. 
The sum in Eq.~(\ref{eq:osci_str}) runs over all considered vertical transitions from valence band $v$ to conduction $c$ at k-point $\mathbf{k}$.
The quantity $F_{vc\mathbf{k}}^{i} = (f_{c\mathbf{k}} - f_{v\mathbf{k}})M^{i}_{vc\mathbf{k}}/(\epsilon_{\nu\mathbf{k}}-\epsilon_{\mu\mathbf{k}})$ includes the transition matrix elements $M^i_{vc\mathbf{k}}=\bra{c\mathbf{k}}\hat{v}_i\ket{v\mathbf{k}}$ of the velocity operator $\hat{v}$ along the Cartesian direction $i$, the Fermi occupation function $f_{n\mathbf{k}}$, and the Kohn-Sham states $\ket{n\mathbf{k}}$ and eigenvalues $\epsilon_{n\mathbf{k}}$ ($n = v,c$) obtained from density functional theory (DFT) calculations.

In Eq.~(\ref{eq:osci_str}) we can see that for each exciton $\Lambda$ all considered single-particle transition matrix elements $M^i_{vc\mathbf{k}}$ contribute with a weight corresponding to the associated normalized BSE eigenvectors $A^\Lambda_{vc\mathbf{k}}$.
This complicates the assignment of spectral features to specific states considerably.

For this reason, fatband analyses of excitonic transitions are often employed in studies of bulk and two-dimensional materials (see, e.g., Fig.~2 in Ref.~\cite{Acharya2023} and Fig.~3 in Ref.~\cite{Varrassi2024}).
In these analyses, the "fatness" (e.g., band thickness or marker size) of the valence and conduction bands is proportional to $\vert A_{vc\mathbf{k}}^\Lambda\vert^2$. 
In practice, one first selects a particular exciton state of interest, i.e., $\tilde{\Lambda}$, and calculates $\vert A_{vc\mathbf{k}}^{\tilde{\Lambda}}\vert^2$ for all transitions included in the BSE Hamiltonian. 
Then, for each transition (indexed by $v$, $c$, and $\mathbf{k}$), the corresponding valence and conduction bands in the band structure are "highlighted" proportional to $\vert A_{vc\mathbf{k}}^{\tilde{\Lambda}}\vert^2$.
The highlighted Kohn-Sham (KS) states can then be assigned to individual atoms or specific orbitals by projecting them onto local orbitals, providing valuable insight into the structure of individual excitons.
However, analyzing RA spectra in this way becomes impractical because, as will be seen later, many excitons often contribute to a specific spectral feature, rendering such an approach cumbersome and inefficient.

For this reason, we introduce the layer-resolved exciton localization (LREL) measure which is calculated for the valence bands (VB) as
\begin{equation}
    \label{eq:lrel}
    \mathrm{LREL}_{\Lambda,\alpha}^\mathrm{VB} = \sum_{vc\mathbf{k}} \vert A^\Lambda_{vc\mathbf{k}}\vert^2 P_{v\mathbf{k}}^\alpha
\end{equation}
where $P_{n\mathbf{k}}^\alpha$ is the layer-resolved projected wave function character of the KS state $\ket{n\mathbf{k}}$.
Here the index $\alpha$ refers to a particular layer in the slab, not to be confused with the half-slab polarizability $\alpha^\mathrm{hs}(\omega)$.
For conduction bands (CB),  $\mathrm{LREL}_{\Lambda,\alpha}^\mathrm{CB}$ is defined analogously with $P_{v\mathbf{k}}^\alpha$ replaced by $P_{c\mathbf{k}}^\alpha$.
Clearly, LREL quantifies the contribution of different layers in a slab to a given exciton.
This is done by weighting the contributions of each layer $\alpha$ to a state $\ket{n\mathbf{k}}$, i.e., $P_{n\mathbf{k}}^\alpha$, by how much that state contributes to excitons $\Lambda$, i.e., $\vert A^\Lambda_{vc\mathbf{k}}\vert^2$, for all single-particle transitions considered.

Similar to the fatband analysis briefly described above, LREL now allows us to quantify how much each layer $\alpha$ of a surface slab contributes to the exciton $\Lambda$ at energy $E^\Lambda$.
Since it involves only reciprocal space quantities, it is easy to compute even for many excitons without having to use, for example, high-dimensional excitonic wave functions on a real-space grid.

In this study, we calculate $P_{n\mathbf{k}}^\alpha$ as follows: 
(i) We calculate the orbital- and site-projected partial wavefunction character of the KS state $\ket{n\mathbf{k}}$, i.e., $C_{lm, n\mathbf{k}}^{\beta}= \vert\langle{Y_{lm}^\beta}\vert{n\mathbf{k}}\rangle\vert^2$ with spherical harmonics $Y_{lm}^\beta$ centered on the ion $\beta$, as implemented in the Vienna Ab-initio Simulation Package (\textsc{VASP}) \cite{Kresse1996, Kresse1999}.
(ii) For each site $\beta$, the sum over all orbitals $\sum_{lm} C_{lm, n\mathbf{k}}^{\beta}$ is calculated.
(iii) By summing the contributions of the individual atoms $\beta$ within the layer $\alpha$ of the surface slab, we obtain $P_{n\mathbf{k}}^\alpha$.
The top layer (hereafter referred to as S) corresponds to the surface reconstruction and consists of either As-As or As-Si-H dimers, depending on the reconstruction, while the other layers consist of two silicon atoms each.

It is important to keep in mind that the numerical values of the projections implicitly depend on several parameters, most notably the projection radii around the atomic sites.
Here, the values set given by the projector augmented wave pseudopotentials \cite{Bloechl1994} for each atomic species were used as projection radii, thereby ensuring consistency with the basis functions and maintaining accuracy without the need for manual adjustments.
For any reasonable choice of projection radii, the LREL should be sufficiently accurate to allow quantitative analysis of the contributions of the different layers and thus insight into the spatial localization of excitonic effects without the need for more complex projection methods.
To make it easier to interpret LREL for a fixed exciton $\Lambda$, we introduce the normalized LREL
\begin{equation}
    \label{eq:nlrel}
    \mathrm{NLREL}_{\Lambda,\alpha}^\mathrm{P} = \frac{\mathrm{LREL}_{\Lambda,\alpha}^\mathrm{P}}{\sum_\alpha \mathrm{LREL}_{\Lambda,\alpha}^\mathrm{P}}
\end{equation}
where $\mathrm{P}=\{\mathrm{VB},\, \mathrm{CB}\}$.
$\mathrm{NLREL}_{\Lambda,\alpha}^\mathrm{VB}$ is now directly proportional to the percentage of how much layer $\alpha$ contributes to the exciton $\Lambda$.
The calculation settings used to obtain $C_{lm, n\mathbf{k}}^{\beta}$ are given in the Methods Section.
Before we present the results, we note two almost trivial observations in passing: 
(i) Any other method of projecting a KS state onto localized orbitals, i.e., obtaining $P_{n\mathbf{k}}^\alpha$, can be used to calculate the LREL. 
(ii) The LREL can also be used to analyze RAS features in an atom- or orbital-resolved fashion by omitting the sum over all atoms in a layer or orbitals. 

\section{Results}

\subsection{Electronic structure and optical properties}

\begin{figure*}[ht]
    \centering
    \includegraphics{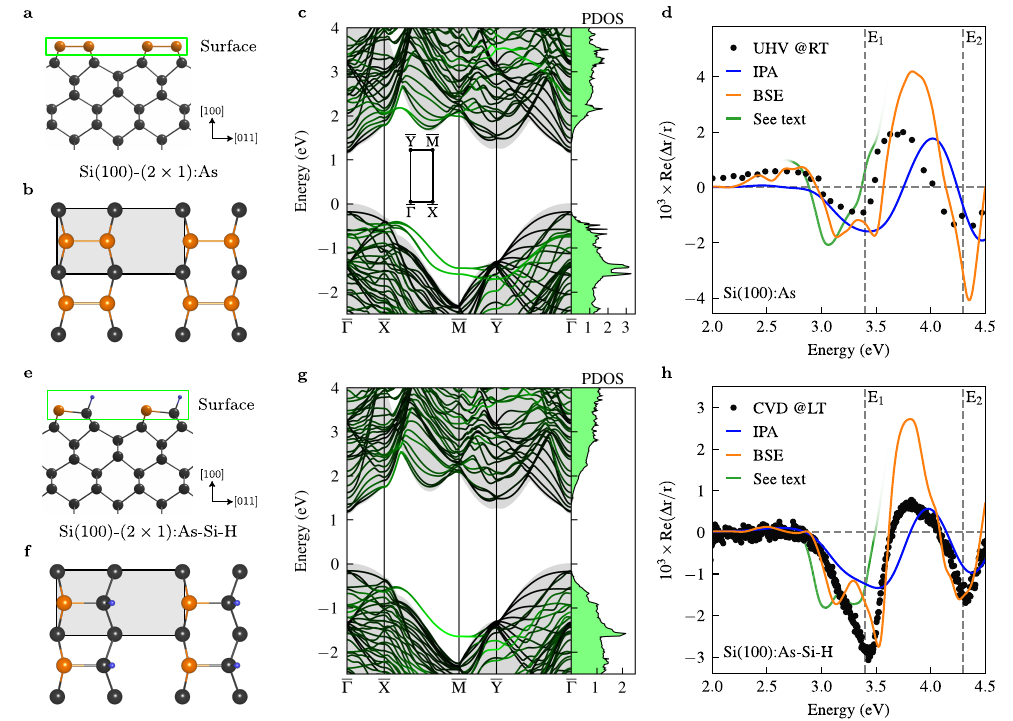}
    \caption{
    Structural, electronic, and optical properties of the Si(100):As (top row, panels \textbf{a}--\textbf{d}) and Si(100):As–Si–H (bottom row, panels \textbf{e}--\textbf{h}) surface reconstructions.
    \textbf{a}, \textbf{e} Side view of the atomic arrangement of the surface reconstructions.
    \textbf{b}, \textbf{f} Top view along the $[100]$ axis of the surface reconstructions. 
    The $(2\times 1)$ surface unit cell is highlighted in gray.
    \textbf{c}, \textbf{g} Surface band structures calculated using the HSE06 functional for 20-layer slabs and partial density of states (PDOS) projected onto the surface atoms highlighted in \textbf{a} and \textbf{e}, respectively. 
    The color of the bands corresponds to $P_{n\mathbf{k}}^\alpha$ for the highlighted surface layer in \textbf{a} and \textbf{e}, where green indicates a high surface contribution and black indicates no surface contribution.
    The projected band structure of bulk silicon is shaded light gray in the background. 
    The $(2\times 1)$ surface Brillouin zone is shown in \textbf{c}.
    \textbf{d} Comparison of RA spectra obtained from experimental measurements and theoretical calculations for the Si(100):As surface reconstruction.
    Experimental data measured at room temperature (RT) have been adapted from Kipp~\textit{et al.}~\cite{Kipp1996}.
    The theoretical RA spectrum with the blue line represents a calculation within the independent particle approximation (IPA) and the orange line representing spectra obtained by solving the Bethe-Salpeter equation (BSE), see text and Methods section.
    The critical points of the imaginary part of the bulk dielectric function of silicon are marked by vertical dashed gray lines \cite{Lautenschlager1987}, highlighting SIBA features.
    \textbf{h} Same as panel \textbf{d}, but for the Si(100):As–Si–H surface reconstruction.
    The experimental RA spectrum was measured by us at low temperatures (LT), see text and the Methods section.
    }
    \label{fig2}
\end{figure*}

\begin{figure*}[ht]
    \centering
    \includegraphics{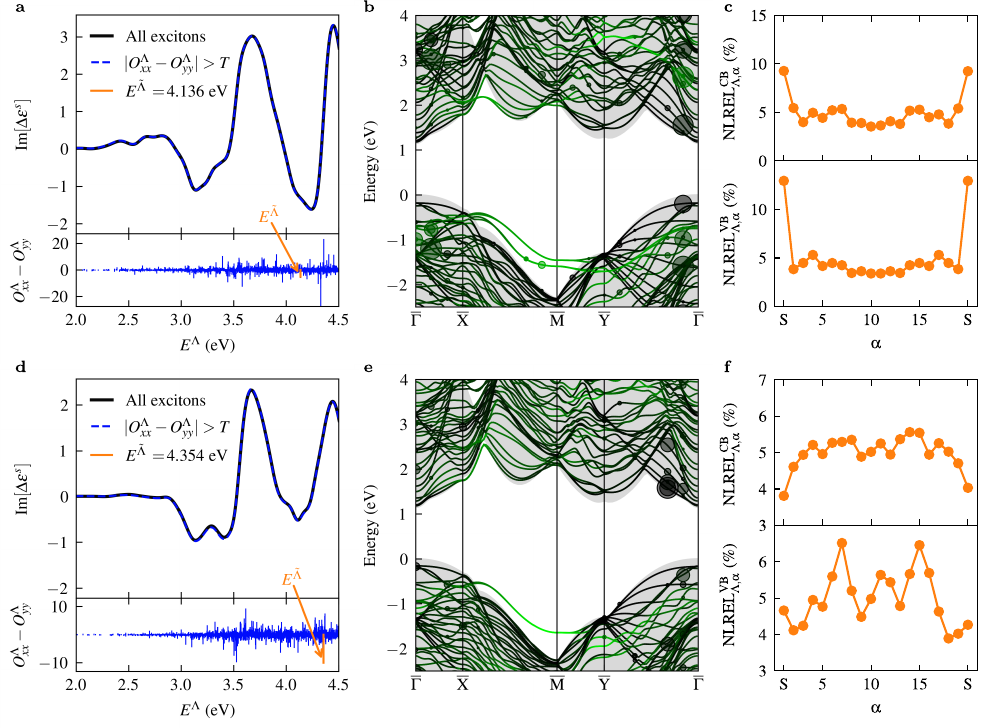}
    \caption{
    Illustration of individual excitons contributing to the surface dielectric anisotropy.
    The top row (\textbf{a}--\textbf{c}) corresponds to the Si(100):As surface reconstruction, while the bottom row (\textbf{d}--\textbf{f}) shows results for the Si(100):As-Si-H surface reconstruction.
    \textbf{a}, \textbf{d} Upper panel: Surface dielectric anisotropy $\Delta \varepsilon^s$ calculated using all excitons (solid black line) and only those with a significant contribution to $\Delta \varepsilon^s$ (dashed blue line), see text. 
    Lower panel: Anisotropy of oscillator strengths for excitons that significantly contribute to $\Delta \varepsilon^s$.
    A particular exciton $\tilde{\Lambda}$ is highlighted in orange (see arrow) and further analyzed in the following columns.
    \textbf{b}, \textbf{e} Fatband analysis of the exciton $\tilde{\Lambda}$ highlighted in orange in the lower panels of \textbf{a} and \textbf{d}, respectively.
    The underlying surface band structures have already been shown and explained in Fig.~\ref{fig2}c and g.
    The size and color of the circles correspond to $\vert A_{vc\mathbf{k}}^{\Lambda} \vert^2$ and $P_{n\mathbf{k}}^\alpha$, respectively.
    Large radii indicate large contributions of the particular KS state to exciton $\tilde{\Lambda}$, and a green color indicates a high surface contribution, while black indicates none.
    The $(2\times 1)$ surface Brillouin zone is shown in Fig.~\ref{fig2}\textbf{c}.
    \textbf{c}, \textbf{f} Normalized LREL measure for the VB and CB for the excitons $\tilde{\Lambda}$ shown in orange in the lower panel of \textbf{a} and \textbf{d} as function of layer index $\alpha$ of the symmetric slab.
    "S" indicates the surface layers.
    }
    \label{fig3}
\end{figure*}

We begin by summarizing the structural, electronic, and optical properties of the Si(100):As and Si(100):As-Si-H surface reconstructions which we need for the subsequent excitonic analysis.
Figure~\ref{fig2} provides an overview of the atomic arrangements, surface band structures, and RA spectra.
The side and top views of the Si(100):As and Si(100):As-Si-H surface reconstructions, highlighting their similar yet distinct atomic configurations, are shown in Fig.~\ref{fig2}a--b and Fig.~\ref{fig2}e--f, respectively.
Surface band structures calculated with the hybrid functional HSE06 (see Methods) are shown in Fig.~\ref{fig2}c and g, where the bands are colored by $P_{n\mathbf{k}}^\alpha$ to highlight the contributions from the surface (green). 
Both reconstructions exhibit surface states near the valence band manifold between the $\overline{\mathrm{X}}$ and $\overline{\mathrm{M}}$ points in the $(2\times 1)$ surface Brillouin zone.
Comparing both, it can be observed that the surface states are directly related to the arsenic atom in the surface structure, as the Si(100):As with two As atoms shows two distinct surface bands, while the Si(100):As-Si-H with only one As atom shows one surface band.
This observation is further supported by the Si(100):As-Si-H surface band structure shown in Supplementary~Note~2, where the bands are colored by $\sum_{lm}C_{lm, n\mathbf{k}}^{\mathrm{As}}$.

In Fig.~\ref{fig2}d and h, we present the experimental and calculated RA spectra.
The measured spectra of both reconstructions exhibit pronounced features at $\mathrm{E}_1$ and $\mathrm{E}_2$ that are also observed in measurements of other Si(100) reconstructions \cite{Palummo2009}. 
The origin of these features, often explained in terms of SIBA, will be investigated later.
Comparing the measured RAS of the Si(100):As at RT with our LT RAS measurements of the Si(100):As-Si-H, we observe a slight blue shift and an increase in the feature amplitudes. 
This observation is more apparent when comparing the Si(100):As-Si-H measured at process temperatures with the LT measurements shown in Supplementary~Note~6.
The observed spectral changes in our LT measurements can be explained as follows:
While thermal expansion causes only a small redshift of the indirect band gap of about $2.5$~meV shift when the temperature is increased from $0$~K to $400$~K in bulk silicon, electron-phonon interactions induce a substantial increase of the indirect band gap at LT of $0.05$--$0.1$~eV \cite{Noffsinger2012}, explaining the shift in peak positions.
The reduction of exciton-phonon scattering at lower temperatures, confirmed by \textit{ab initio} calculations for silicon and h-BN \cite{Marini2008}, leads to the observed reduced peak broadening, i.e., the increase of feature amplitudes. 
In addition, the increase in features amplitude in LT measurement is also likely enhanced by a higher surface order, increasing the anisotropy. 

Having presented the experimental data, we now turn to the spectra obtained from the theory.
The spectra shown are calculated using Eq.~(\ref{eq:ras_ai}), where $\varepsilon^b$ and $\Delta \alpha^{hs}$ are obtained from consistent theory levels starting from a HSE06 calculation, i.e., by calculating the IPA dielectric function or solving the BSE. 
As shown several times before \cite{Rohlfing1999, Hahn2001, Schmidt2003, Palummo2009}, we again observe that excitonic effects included in the BSE (orange) strongly modulate (red shift and line shape change) the RA spectra compared to IPA calculations (blue).
The calculated BSE spectra closely match the experimental data. 
However, it could be argued that the BSE prediction for the Si(100):As surface reconstruction agrees with the measurement by Kipp~\textit{et al.}~\cite{Kipp1996} not much better than the IPA spectra.
This is likely due to the fact that the measurement was performed at RT, where excitonic effects are diminished and some degree of surface disorder may be present, reducing the optical anisotropy as discussed above.
Comparing the two surface reconstructions, we notice a reversal in the amplitude ratio of the peaks at $\mathrm{E}_1$ and $\mathrm{E}_2$, a trend also suggested by experimental data.
However, a LT measurement of Si(100):As would be required to fully validate this theoretical observation.
At this point, we need to comment on the double peak visible at $\mathrm{E}_1$ in both BSE calculations (orange lines in Fig.~\ref{fig2}d and h). 
A similar feature has been observed and extensively discussed by Albrecht~\textit{et al.}~\cite{Albrecht1999Comment} in the case of calculated dielectric function of bulk silicon, where it was referred to as a 'k-point ripple', i.e., a numerical artifact resulting from insufficiently dense k-point sampling.
To check whether the double peak visible at $\mathrm{E}_1$ in the RA spectra might be due to a similar convergence problem, we performed additional calculations (green curves) using a k-point grid twice as dense as that used for the orange curves. 
Due to computational limitations, these additional calculations were performed with a reduced transition space covering an energy range just above $3.5$~eV and were based on scissored PBEsol energies and wavefunctions rather than the more computationally demanding HSE06 approach.
Comparing the green and orange curves in Fig.~\ref{fig3} we find that the double peak at $\mathrm{E}_1$ slowly merges into a single feature when the k-point sampling is doubled, confirming that the double peak is indeed not a feature but a 'k-point ripple'.
However, solving the BSE utilizing energies and wavefunction from a lower level of theory (scissored PBEsol) shifts the predicted excitations significantly, see green curves.
We have therefore decided to continue using the more accurate HSE06 approach in all main calculations, despite the slightly underconverged k-point sampling due to computational limitations, in order to obtain an overall higher level theoretical description.

\begin{figure*}[ht]
    \centering
    \includegraphics{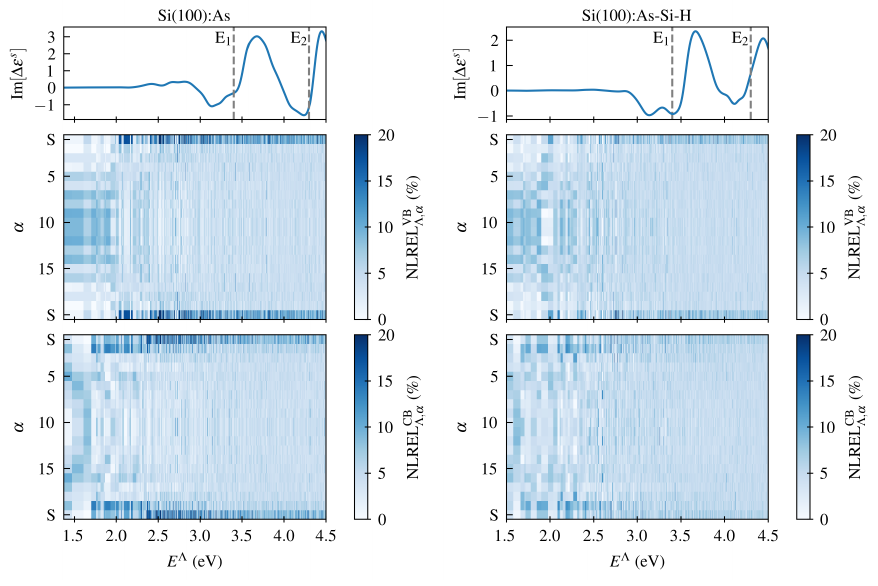}
    \caption{
    Illustration of the surface dielectric anisotropy (upper panels) and NLREL maps calculated using Eqs.~(\ref{eq:lrel}--\ref{eq:nlrel}) (lower panels) for the VB and CB manifolds, respectively.
    Left column: Si(100):As surface reconstruction.
    Right column: Si(100):As-Si-H surface reconstruction.
    In the upper panels, the critical points of the imaginary part of the bulk dielectric function of silicon are marked by vertical dashed gray lines \cite{Lautenschlager1987}.
    We limit both color bars to $20$\% to facilitate comparison between the two reconstructions, as few excitons in the Si(100):As surface reconstruction have layer localizations above this threshold.
    }
    \label{fig4}
\end{figure*}

\subsection{Exciton analysis}

As demonstrated by our toy model in the introduction, analysis of RA spectra alone provides limited insight into the origin of RAS features, especially SIBA features, as BESA effects may shroud their origin (cf.~Fig.~\ref{fig1}). 
To address this, we focus directly on the surface dielectric anisotropy $\Delta \varepsilon^s$.
To simplify the analysis, we consider only excitons that contribute strongly to the RAS, defined as those with $\vert O^\Lambda_{xx} - O^\Lambda_{yy} \vert > \mathcal{T}$, where $\mathcal{T}$ is a threshold set to the average contribution of all excitons within the RAS energy range (see Supplementary~Note~3 for details). 
Even with this constraint, approximately $10^4$ excitons remain for analysis for each surface reconstruction. 
In the upper panels of Fig.~\ref{fig3}a and d, we show the surface dielectric anisotropy for both reconstructions, demonstrating that our exciton subset already accurately reproduces the surface dielectric anisotropy. 
Furthermore, the lower panels of Fig.~\ref{fig3}a and d show that individual excitons do not dominate the contributions to each RAS feature. 
Instead, each feature results from a complex summation of positive and negative contributions, with the dominant contribution emerging only when all excitons are considered together.
Note that the imaginary part of the surface dielectric anisotropies already looks remarkably similar to the RA spectra shown in Fig.~\ref{fig2}d and h, with negative anisotropies at $\mathrm{E}_1$ and $\mathrm{E}_2$.
Therefore, we conclude that the SIBA features seen in the RAS in Fig.~\ref{fig2}d and h are not only caused by the BESA effect visible in the model system (cf.~Fig.~\ref{fig1}).
To investigate the origin of these features in the imaginary part of the surface dielectric function, we now focus on a detailed analysis of the contributing excitons.

First, we illustrate how a conventional fatband analysis and our NLREL measure can be used to analyze the origin of individual excitons. 
To do this, we selected one exciton $\tilde{\Lambda}$ for each surface reconstruction, highlighted in orange in the bottom panels of Fig.~\ref{fig3}a and d.
Figures~\ref{fig3}b and \ref{fig3}e show the corresponding surface band structures, with the projection $P_{n\mathbf{k}}^\alpha$ onto the surface layer highlighted in green (cf.~Fig.~\ref{fig2}d and g). 
We have added circles to the surface band structures for all transitions contributing to the oscillator strength, whose size is proportional to $\vert A_{vc\mathbf{k}}^\Lambda \vert^2$, while their color indicates the projection: black circles correspond to bulk states, and green circles represent states localized on surface atoms.
The NLREL values for the selected excitons are shown in Fig.~\ref{fig3}c and f. 

For the Si(100):As reconstruction (Fig.~\ref{fig3} top row) we have chosen an exciton with a strong contribution from surface states. 
The fatband representation of the surface band structure shown in Fig.~\ref{fig3}b, highlights the localization of this particular exciton at surface states throughout the whole surface Brillouin zone (green circles), with some contributions from bulk states between to $\overline{\mathrm{Y}}$ and $\overline{\Gamma}$ (black circles). 
This observation is directly supported by the NLREL plot in Fig.~\ref{fig3}c. 
The VB and CB contributions are localized directly at the surface, while all other layers contribute uniformly.

For the Si(100):As-Si-H reconstruction (Fig.~\ref{fig3} bottom row) we have chosen an exciton with strong bulk contributions. 
The fatband representation of the surface band structure in Fig.~\ref{fig3}e shows much more pronounced bulk transitions between the $\overline{\mathrm{Y}}$ and $\overline{\Gamma}$ points, visible as black circles.
Again, the NLREL in Fig.~\ref{fig3}f reflects this, showing that the exciton is more localized in the center of the slab compared to the surface layer, with layer contributions increasing as one goes deeper into the slab.
Here one should not be surprised that the NLREL in Fig.~\ref{fig4}f is not symmetric around the center of the slab, since the slab has only a center of inversion and not perfect mirror symmetry around the middle layer. 

In summary, one could use the presented fatband analysis in conjunction with our NLREL to study the localization of individual excitons. 
However, this is impractical for the subset of $10^4$ excitons that all contribute to the RAS in some non-trivial way (see the lower panels of Fig.~\ref{fig3}a and \ref{fig3}d), so we resort to the NLREL maps shown in Fig.~\ref{fig4}, which streamline the analysis considerably.

However, before discussing the actual NLREL maps shown in Fig.~\ref{fig4}, it is worth contemplating what one would expect these maps to look like, especially around the SIBA features at $\mathrm{E}_1$ and $\mathrm{E}_2$.
Recalling the quote about the origin of the SIBA features, i.e., "bulk-like electronic states perturbed by the surface" from Ref.~\cite{Schmidt2000} and the conclusion drawn from Ref.~\cite{Hahn2001}: "It is shown that excitonic effects via strong modifications of the optical response of surface-modified bulk wave functions determine largely the line shape of the optical features.", we would expect the following:
Features in the surface dielectric anisotropy shown in Fig.~\ref{fig3}a and d at $\mathrm{E}_1$ and $\mathrm{E}_2$ would be localized in layers deeper in the slab toward the 'bulk' region, caused by perturbations of the bulk wave function by the surface reconstructions. 
Other features may originate directly from the surface, i.e., the NLREL map would show strong localization at or just below the surface. 

Looking at the NLREL maps in Fig.~\ref{fig4}, our expectations are not met.
In the Si(100):As surface reconstruction, almost all of the excitons above $2$~eV are localized at the surface.
The few excitons below $2$~eV are localized in the 'bulk' region of the slab, while having almost no anisotropy, as can be seen by looking at the surface dielectric anisotropy in the upper panel of the left column of Fig.~\ref{fig4}.
The VB and CB contribution looks similar. 
For Si(100):As-Si-H, the VB states of the excitons below $2.5$~eV are similar to those of the Si(100):As reconstruction. 
However, the CB states are localized in the first layer below the surface.
Again, these excitons have a vanishing contribution from the surface dielectric anisotropy.
When comparing the NLREL maps for both surfaces above $3$~eV, differences become apparent, especially when comparing the CB states of the excitons.
In the Si(100):As-Si-H surface reconstruction, the contribution of all layers to the excitons is averaged, and no strong surface or bulk localization is visible.
However, the VB states are more localized at the surface, while all other layers have a small and similar contribution, similar to Si(100):As.
After comparing the NLREL maps of both surfaces, it is clear that the excitons of the symmetric dimer motif of the Si(100):As reconstruction are much more localized at the surface compared to the asymmetric Si(100):As-Si-H reconstruction, yet their surface dielectric anisotropies are remarkably similar.
In both systems we also find that the surface contribution tends to weaken slightly as one goes to higher energies, approaching the $\mathrm{E}_2$ peak.

\begin{figure}[ht]
    \centering
    \includegraphics{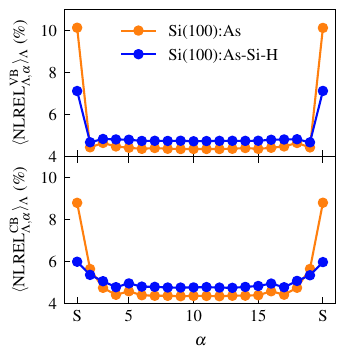}
    \caption{Energy averages of the NLREL maps shown in Fig.~\ref{fig4}.}
    \label{fig5}
\end{figure}

\begin{figure*}[ht]
    \centering
    \includegraphics{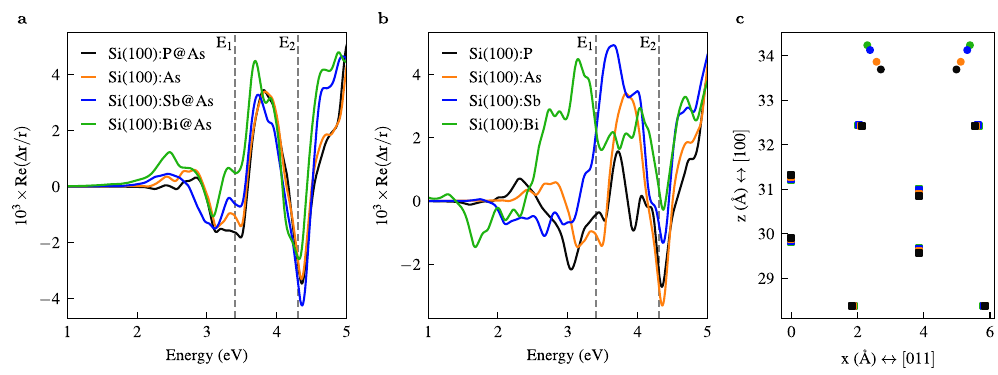}
    \caption{
    \textbf{a} RA spectra for Si(100):V@As surfaces calculated with the fixed Si(100):As geometry.
    \textbf{b} RA spectra for the relaxed Si(100):V surfaces.
    \textbf{c} Structural view (of the $xz$ plane spanned by $[011]$ and $[100]$ vectors) of the top layers of the relaxed Si(100):V systems.
    Square markers denote Si atoms, while circular markers denote group V atoms, as indicated by the color coding, which is consistent across all panels.
    }
    \label{fig6}
\end{figure*}

Now that we have discussed the NLREL maps and highlighted their differences, we question the term 'SIBA' commonly used to label features that coincide with the critical energies of the bulk dielectric function, since strong bulk contributions are missing from both 'SIBA' features in the NLREL maps, at least for Si(100):As.
One could argue, however, that in the case of the Si(100):As-Si-H reconstruction, the features in the surface dielectric anisotropy around $\mathrm{E}_1$ and $\mathrm{E}_2$ may still be labeled 'SIBA' because many bulk-like states from many layers in the slab contribute to the surface dielectric anisotropy.
No clear surface localization is visible, yet no clear localization of excitons in the 'bulk' region of the slab is visible either, as would be expected, similar to the VB states of the excitons below $2.5$~eV. 
The previously proposed picture of "bulk-like electronic states perturbed by the surface" \cite{Schmidt2000} may still be correct, but in the broader sense that all states, independent of the layer, are perturbed by the surface.
Looking at the Si(100):As surface, the general labeling of features that coincide with the critical energies of the bulk dielectric function as 'SIBA' breaks down completely because most of the excitons contributing to the surface dielectric anisotropy are localized at the surface.
Both of the above observations become even clearer when looking at the energy averaged NLREL in Fig.~\ref{fig5}.
Here the difference between the surface contributions for the Si(100):As-Si-H and all other layers is small.
For the Si(100):As the difference is significant.

As a side note, it would be interesting to know how these NLREL maps would look if the layer size were increased drastically.
At the moment we are computationally limited in this regard, but we have checked NLREL maps with 10 layers using the same calculation parameters and observed very similar results.

\subsection{Deciphering optical anisotropy of Si(100):V surfaces}

Having shown that the NLREL maps can be used to identify the spatial origin of features in surface dielectric anisotropies, and that the origin of 'SIBA' features is strongly system dependent, the physical effects behind them remain elusive.
In particular, the cause of the strong surface localization of excitons above $2$~eV in the Si(100):As deserves further investigation.
To gain insight into the mechanism behind this localization, we performed further calculations on Si(100)-$(2\times 1)$:V surface reconstructions with symmetric dimers.
Specifically, we repeated all the calculations presented for six additional systems. 
For three of them, we kept the Si(100):As geometry and replaced the As atoms with other group V atoms, i.e., P, Sb, and Bi.
We will refer to these systems as Si(100):V@As (V = P, Sb, Bi) and again omit the $(2\times 1)$ from the surface reconstruction notation.
The pseudopotentials for all group V atoms have the same valence electron configuration and differ only in the treatment of their core electrons and atomic size.
The other three systems also follow the motif of a Si(100) surface reconstruction with a symmetric dimer group V-dimer in a $(2\times 1)$ supercell, but this time we have relaxed their geometry and will refer to them as Si(100):V.
The calculations for six additional surfaces are performed with the same parameters as the previous ones for consistency, see Methods.

First, we comment on the Si(100):V@As surfaces calculated with a fixed Si(100):As geometry.
Looking at their RA spectra in Fig.~\ref{fig6}a, we observe that they all look remarkably similar, and interestingly, the $\mathrm{E}_1$ anisotropies follow the order of the electronegativities of the group V elements, suggesting at least a small contribution of electric-field-induced anisotropy to the RAS \cite{AcostaOrtiz1989, WGS-E-Feld}.
The associated surface band structures, the surface-projected density of states, and the NLREL maps shown in Supplementary~Note~4 are also quite similar.
In particular, the surface band structures show a strong similarity, with some differences in the gap between the two surface states propagating the bulk band gap close to the valence bands between the $\overline{\mathrm{X}}$ and $\overline{\mathrm{M}}$ points.
This may be explained by the following simple argument: 
The geometry of the Sb and Bi dimers is drastically compressed (cf.~Fig.~\ref{fig6}c), leading to an increase in the 'band gap' of their associated surface states, similar to the expected increase in band gap when a solid is compressed.
This causes the surface states to look similar to those of an As dimer.
In the case of the P dimer, the inverse argument may be applied.

Looking at the RA spectra in Fig.~\ref{fig6}b for the relaxed Si(100):V geometries shown in Fig.~\ref{fig6}c, we observe a strong change in the spectra below the $\mathrm{E}_2$ feature.
This is in line with the model calculations shown in Fig.~\ref{fig1}:
Now that the dimers are relaxed (decrease in P-P bond distance, increase in Sb-Sb and Bi-Bi bond distances, cf.~Fig.~\ref{fig6}c), the 'band gap' between the associated surface states shrinks. 
Indeed, as shown in Supplementary~Note~4, the surface bands shift toward the bulk band gap throughout the 2D Brillouin zone.
This changes the transition space and thus the excitonic structure, which can be observed as features at lower energies in the RA spectra around and below $\mathrm{E}_1$.
This enhanced contribution of surface states to the RA spectra and surface dielectric anisotropy is further highlighted by the corresponding NLREL maps in Supplementary~Note~4.

Comparing the RAS (and additional illustrations in Supplementary~Note~4) of the constraint and relaxed Si(100):V surfaces reveals the following:
(i) Surface geometry is clearly the most dominant influence on RA spectra as it affects all parts of the electronic and derived optical properties. 
This even concerns the localization of the excitons that contribute most to the optical anisotropies.
(ii) The $\mathrm{E}_1$ feature in the RA spectrum of Si(100):As is coincidentally at the $\mathrm{E}_1$ energy of bulk silicon and has nothing to do with 'SIBA', i.e., "bulk-like electronic states perturbed by the surface" \cite{Schmidt2000}, as it originates from surface states.
This is further emphasized by the fact that the shape and position of this feature changes drastically in all other relaxed Si(100):V surfaces. 
(iii) The $\mathrm{E}_2$ feature is clearly visible as a pronounced peak in all RA spectra (and surface dielectric anisotropies, see Supplementary~Note~4). 
We therefore argue that this feature is a mixture of some surface contributions causing intensity changes and contributions from deeper layers, since surface contributions tend to weaken at higher energies around $\mathrm{E}_2$ in all NLREL maps (see Supplementary~Note~4).
In addition, the BESA effect highlighted in the simple model system, cf.~Fig.~\ref{fig1} and associated text, also contributes to the presence of the $\mathrm{E}_2$ feature in all systems.

\section{Discussion}

Our results provide new insight into the long-standing debate on the origin of bulk-related anisotropies in surface optical spectra. 
Traditionally, such features have been attributed to SIBA, i.e., "bulk-like electronic states perturbed by the surface" \cite{Schmidt2000}.

First, we use a simple model system to show that these apparent 'SIBA' features may arise from what we call bulk-enhanced surface anisotropy (BESA). 
Obviously, a closer examination of the dielectric anisotropy is needed to gain thorough insight into the origin of these features.
Diving deeper, we perform an \textit{ab initio} analysis of the excitonic contributions to the surface dielectric anisotropy of the technologically relevant arsenic-terminated Si(100) surface reconstruction utilizing the framework of MBPT.
We introduce the LREL measure, which allows us to obtain a layer-resolved map of exciton origins, thereby disentangling the contributions of surface and subsurface states to the RA spectra.
Again, our findings challenge the simple SIBA picture, as excitonic contributions to 'bulk-related' features in RA spectra can be of different origins. 
For example, in the Si(100):As system, most excitons contributing to the optical anisotropy are localized directly at the surface, whereas they are rather delocalized for Si(100):As-Si-H.
In the latter case, one could indeed speak of a SIBA effect.  
The comparison of NLREL results for constraint and relaxed Si(001):V surfaces shows that the surface localization of anisotropy-related excitons, in fact, depends sensitively on the structural details of the surface. 
 
We identify three possible origins of RAS features that were previously attributed to SIBA: (i) BESA effects, where the bulk dielectric function strongly modulates the optical anisotropy, as illustrated in Fig.~\ref{fig1}; (ii) features from surface states that coincidentally appear at bulk energies, as clearly seen in the Si(100):As reconstruction; and (iii) SIBA effects in the true sense of the word, as observed, e.g., for Si(100):As-Si-H. 
The present findings thus underline that a general classification of optical anisotropies at bulk critical points energies as SIBA does not take into account the complex origin of surface optical spectra.
Rather, a detailed exciton-resolved analysis is indispensable for a proper understanding of the spectra.

The generality and simplicity of our LREL method makes it readily applicable to other semiconductor surfaces and heterostructures.
By directly quantifying the contributions from individual atomic layers, our approach allows to disentangles the relative roles of surface and subsurface states in optical spectra and provides a quantitative framework that can be used to reexamine anisotropy features in previously studied systems. 

In summary, the present work shows that 'bulk-related' anisotropies in surface optical spectra do not necessarily originate from the bulk, but may as well be bulk-enhanced surface anisotropies, challenging the conventional SIBA interpretation.
By introducing a versatile, general methodology for analyzing exciton localization, our work paves the way for a refined interpretation of surface optical responses that fully accounts for the complex interplay between surface chemistry, excitonic effects, and bulk contributions.
We anticipate that these findings will spur further experimental and theoretical efforts to unravel the intricacies of surface optical phenomena.

\section{Methods}\label{methods}

\subsection*{Ab initio calculations}

DFT and MBPT calculations were performed with a modified version of the version of \textsc{VASP} 5.4.4 \cite{Kresse1996, Kresse1999}, which uses a plane-wave basis and the projector augmented-wave method \cite{Bloechl1994}.
Details about the code modification can be found in Supplementary~Note~3.
All key calculation parameters (plane-wave cutoff, k-point grid, vacuum size, and slab thickness) were systematically converged so that the ground state energy---and the surface energy for vacuum size and slab thickness---varied by less than $1$~meV. 
A plane-wave cutoff of $400$~eV and the PBEsol functional \cite{Perdew2008, Perdew2009} were used for all calculations unless otherwise noted.
PBEsol was chosen, in particular to obtain accurate bulk and surface geometries.
For all systems, the lattice constant was set to the value calculated with PBEsol, i.e., the Si bulk lattice constant of $5.436$~{\AA}.
The surfaces were modeled as non-vicinal, 20-layer slabs \cite{DelSole1999} with a center of inversion separated by a 16~{\AA} vacuum and relaxed until the norms of all forces were smaller than $10^{-3}$~eV/{\AA}.
Unless otherwise noted, surface calculations used a $6 \times 12 \times 1$ $\Gamma$-centered k-point grid, while bulk calculations used a $12 \times 12 \times 12$ grid. 
Due to the high computational cost of GW calculations for surfaces, including the associated calculations needed to find converged parameters \cite{Grossmann2024} and the necessary treatment of long-range screening \cite{Freysoldt2008}, we resorted to the HSE06 functional \cite{Heyd2003,Heyd2006} to obtain a good approximation of the quasiparticle band structure for our relaxed slab models. 
Starting from the HSE06 results, we solved the BSE \cite{Onida2002} within the Tamm-Dancoff approximation \cite{Bechstedt2014} using an analytical model for static screening \cite{Bechstedt1992} to obtain the surface dielectric tensor $\varepsilon_{ii}^s(\omega)$ and the bulk dielectric function $\varepsilon_b(\omega)$. 
Details and parameters for the static screening model are given in Supplementary~Note~5.
The broadening parameter $\gamma$ from Eq.~(\ref{eq:bse_eps}) has been set to $0.1$~eV.
The BSE Hamiltonian for the surfaces included $36$ occupied and unoccupied Kohn-Sham orbitals.
For the bulk, we included $4$ occupied and $12$ unoccupied Kohn-Sham orbitals on a dense $24 \times 24 \times 24$ $\Gamma$-centered k-point grid.
In all BSE calculations we have removed from the basis set valence and conduction band pairs whose difference in independent particle energy is greater than $10$~eV (\texttt{OMEGAMAX}=10).
The orbital- and site-projected partial wavefunction characters $\ket{n\mathbf{k}}$, i.e., $C_{lm, n\mathbf{k}}^{\beta}$, were calculated with \textsc{VASP} input tags \texttt{LORBIT=11} and \texttt{RWIGS=-1}, i.e., the values set in the PAW pseudopotentials were used as projection radii.
The convergence of the BSE results with respect to the k-point grid is discussed in the main text.

\subsection{Sample preparation and measurements}

We used $p$-doped Si(100) substrates with a $6^{\circ}$ offcut angle towards the $[011]$ direction to prepare an A-type \cite{Paszuk2018b} Si(100)-$(1\times 2)$:As-Si-H surface, analogous to Ref.~\cite{Bohlemann2024}.
Here we want to emphasize that the difference between $(1\times 2)$ and $(2\times 1)$ affects the orientation of the dimers along or perpendicular to the step edges of a vicinal surface, and therefore this is only an "experimental" detail. 
This difference is well illustrated in Fig.~1 of Ref.~\cite{Paszuk2018b}.
However, throughout the rest of the paper, we always refer to a $(2\times 1)$ reconstruction, since all \textit{ab initio} calculations have been performed for non-vicinal surfaces, making this distinction irrelevant.
The samples were prepared using an H$_2$-based horizontal AIX-200 MOCVD reactor (Aixtron).
The entire MOCVD process was monitored \textit{in situ} by RAS (LayTec EpiRAS-200). 
The Si(100) substrates here were first deoxidized by annealing under constant hydrogen flow for $30$~min at $1000^{\circ}\mathrm{C}$ and $950$~mbar in MOCVD.
Subsequently, the surface was prepared by exposure to tertiarybutylarsine (TBAs) at $800^{\circ}\mathrm{C}$ and 950~mbar.
Here, a sequence of opening and closing the TBAs source was applied while maintaining a constant hydrogen flow \cite{Paszuk2018b}. 
This stepwise exposure increases the surface order, and exposure to TBAs prepares double layer steps on the surface, resulting in more pronounced RAS peak amplitudes.
After exposure, the samples were cooled under constant hydrogen flow and RA spectra were recorded at $597$~K as shown in Supplement~Note~6.

After preparation, the samples were first transferred to a cryostat (Montana Instruments Cryostation) using an MOCVD-to-UHV shuttle to prevent surface contamination \cite{Hannappel2004}.
Here, the custom copper cryostat head was first cooled by a closed-loop helium setup that reduced the pressure inside the cryostat from HV to UHV as it cooled. 
The sample, mounted on a molybdenum carrier, was then inserted into the copper cooling head while a series of LT RA spectra were acquired at a sample temperature of approximately $50$~K.
Details regarding sample temperature estimation are given in Supplement~Note~6.
Following the LT RAS measurements, the sample was further transferred to LEED (specs ErLEED 100), XPS (monochromated Al-K$\alpha$, specs Focus 500/Phoibos 150/1D-DLD-43-100) to ensure that no contamination had occurred during sample transfer or in the cryostat that would cause structural or chemical changes, see Supplementary~Note~7.

\section{Data and Code Availability}

The data generated and the post-processing code used in this study are available from the corresponding author upon reasonable request.

\section{Acknowledgments}

We thank the staff of the Compute Center of the Technische Universität Ilmenau and especially Mr.~Henning~Schwanbeck for providing an excellent research environment. 
M.G. would like to thank Malte Grunert for asking thought-provoking questions, providing constructive feedback as a proofreader, and contributing to a delightful and productive office atmosphere.
This work is supported by the Carl Zeiss Stiftung (funding code: P2023-02-008), the German Federal Ministry of Education and Research (BMBF, proj. H2Demo, proj. no. 03SF0619I), and the DFG (TRR 142/3-2025 proj. no. 231447078). 
We acknowledge the Paderborn Center for Parallel Computing (PC$^2$) for grants of high-performance computer time.

\section{Competing interests}

The authors declare no competing interests.

\section*{Author contributions}

M.G. conceived the idea, carried out the calculations, analyzed the data, visualized all results, and wrote the first draft of the manuscript. 
E.R. suggested the model calculations.
K.D.H. prepared the samples, performed the LEED, XPS, and LT RAS measurements and evaluated the experimental results.
C.Y.B. assisted the sample preparation and evaluation of experimental results.
A.P. advised the experimental planning.
M.G., W.G.S., and E.R. discussed and interpreted the theoretical results.
T.H., W.G.S., and E.R. supervised the work; all authors revised and approved the manuscript.


\providecommand{\noopsort}[1]{}\providecommand{\singleletter}[1]{#1}%

\end{document}


\title{Supplementary Information for "On the origin of bulk-related anisotropies in surface optical spectra"}

\author{Max Großmann}
\email{max.grossmann@tu-ilmenau.de}
\affiliation{Theoretical Physics I, Institute of Physics, Technische Universit\"at Ilmenau, 98693 Ilmenau, Germany}
\affiliation{Center of Micro- and Nanotechnologies, Technische Universit\"at Ilmenau, 98693 Ilmenau, Germany}

\author{Kai Daniel Hanke}
\affiliation{Center of Micro- and Nanotechnologies, Technische Universit\"at Ilmenau, 98693 Ilmenau, Germany}
\affiliation{Fundamentals of Energy Materials, Institute of Physics, Technische Universit\"at Ilmenau, 98693 Ilmenau, Germany}

\author{Chris Yannic Bohlemann}
\affiliation{Center of Micro- and Nanotechnologies, Technische Universit\"at Ilmenau, 98693 Ilmenau, Germany}
\affiliation{Fundamentals of Energy Materials, Institute of Physics, Technische Universit\"at Ilmenau, 98693 Ilmenau, Germany}

\author{Agnieszka Paszuk}
\affiliation{Center of Micro- and Nanotechnologies, Technische Universit\"at Ilmenau, 98693 Ilmenau, Germany}
\affiliation{Fundamentals of Energy Materials, Institute of Physics, Technische Universit\"at Ilmenau, 98693 Ilmenau, Germany}
\affiliation{BMBF Junior Research Group PARASOL, Institute of Physics, Technische Universit\"at Ilmenau, 98693 Ilmenau, Germany}

\author{Thomas Hannappel}
\affiliation{Center of Micro- and Nanotechnologies, Technische Universit\"at Ilmenau, 98693 Ilmenau, Germany}
\affiliation{Fundamentals of Energy Materials, Institute of Physics, Technische Universit\"at Ilmenau, 98693 Ilmenau, Germany}

\author{Wolf Gero Schmidt}
\affiliation{Lehrstuhl f\"ur Theoretische Materialphysik, Universit\"at Paderborn, 33095 Paderborn, Germany}

\author{Erich Runge}
\affiliation{Theoretical Physics I, Institute of Physics, Technische Universit\"at Ilmenau, 98693 Ilmenau, Germany}
\affiliation{Center of Micro- and Nanotechnologies, Technische Universit\"at Ilmenau, 98693 Ilmenau, Germany}

\date{\today}

\maketitle

\section*{Supplementary Note 1: Details on the model system}

In the main text, we discussed how the interplay between the bulk dielectric function $ \varepsilon^b(\omega) $ and the surface dielectric anisotropy $ \Delta \varepsilon^s $ affects the reflection anisotropy (RA) spectra. 
In particular, we calculate
\begin{equation}
    \label{eq:ras_split}
    \mathrm{Re}\left[\frac{\Delta r}{r}\right] = \frac{4\pi d \omega}{c}\,\bigl( A\cdot\mathrm{Im}[\Delta\varepsilon^s] - B\cdot\mathrm{Re}[\Delta \varepsilon^s] \bigr),
\end{equation}
with the bulk-related parameters $A$ and $B$ defined as
\begin{align}
\label{eqn:A}
    A &= \frac{\mathrm{Re}[\varepsilon^b] - 1}{\left( \mathrm{Re}[\varepsilon^b]-1 \right)^2 + \left( \mathrm{Im}[\varepsilon^b] \right)^2},\\
\label{eqn:B}
    B &= \frac{\mathrm{Im}[\varepsilon^b]}{\left( \mathrm{Re}[\varepsilon^b]-1 \right)^2 + \left( \mathrm{Im}[\varepsilon^b] \right)^2}.
\end{align}
In our simple model, we used silicon as the bulk (substrate) material.
To obtain $\varepsilon^b(\omega) $ for silicon, we retrieved experimental optical constants measured by Aspnes~\textit{et al.}~\cite{Aspnes1983} from the Refractiveindex.info database \cite{Polyanskiy2024}. 
The experimental data were processed in two steps:
\begin{itemize}
    \item[(i)] Starting from the optical constants, the complex bulk dielectric function was calculated via
    \begin{equation*}
        \varepsilon^b(\omega) = n^2(\omega) - k^2(\omega) + 2\mathrm{i} n(\omega)k(\omega).
    \end{equation*}
    \item[(ii)] To obtain a high-resolution description, the experimental data were linearly interpolated onto a dense energy grid covering $\omega \in [2.0, 6.0]$~eV.
\end{itemize}
Using the interpolated $\varepsilon^b(\omega)$, the parameters $A$ and $B$ have been calculated by Eqs.~(\ref{eqn:A}-\ref{eqn:B}), see Supplementary~Fig.~\ref{fig:ras_model_si}.

\begin{figure}[ht]
    \centering
    \includegraphics{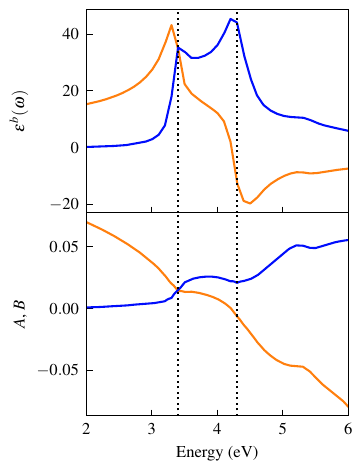}
    \caption{
    Top panel: Real (orange) and imaginary (blue) parts of the bulk dielectric function $\varepsilon^b(\omega)$ for silicon \cite{Aspnes1983}. 
    Bottom panel: Corresponding quantities $A$ (orange) and $B$ (blue) derived from $\varepsilon^b(\omega)$.
    The dashed vertical lines indicate the critical point energies of bulk silicon \cite{Lautenschlager1987}.
    }
    \label{fig:ras_model_si}
\end{figure}

We modeled the surface dielectric anisotropy $\Delta \varepsilon^s$ using a single Lorentz oscillator:
\begin{equation}
    \varepsilon^s(\omega) = \varepsilon_\infty + \frac{A}{\omega_0^2 - \omega^2 - i\,\gamma\,\omega}.
\end{equation}
As already stated in the main text, the anisotropic response is enforced by assuming $ \varepsilon_{yy}^s(\omega) = 0.9\,\varepsilon_{xx}^s(\omega) $, such that $ \Delta \varepsilon = 0.1\,\varepsilon_{xx}^s $. 
In our model calculations, we systematically varied the four oscillator parameters: the resonance energy $\omega_0$, the infinite frequency limit of the dielectric constant $\varepsilon_\infty$, the oscillator strength $A$, and the damping constant $\gamma$. 
When investigating the influence of one parameter, we varied it through a range of reasonable values (see in Supplementary~Table~\ref{tab:params}), while keeping the other three model parameters fixed at their default values, which are highlighted in bold in Supplementary~Table~\ref{tab:params}.
\begin{table}[ht]
    \centering
    \caption{
    Lorentz oscillator model parameters. 
    Default values are highlighted in bold.
    }
    \label{tab:params}
    \begin{tabular}{cccccccc}
        \toprule
        Parameter & 1 & 2 & 3 & 4 & 5 & 6 & 7 \\ 
        \midrule
        $\omega_0$ (eV) & $2.50$ &  $2.95$ &  $3.40$ & $\mathbf{3.85}$ &  $4.30$ & $4.75$ &  $5.20$ \\
        $\varepsilon_\infty$ & $2.0$ & $5.0$ & $7.5$ & $\mathbf{10.0}$ & $12.5$ & $15.0$ & $17.5$ \\
        $A$ & $50$ & $100$ & $150$ & $\mathbf{200}$ & $250$ & $300$ & $350$ \\
        $\gamma$ & $0.25$ & $0.50$ & $0.75$ & $\mathbf{1.00}$ & $1.25$ & $1.50$ & $1.75$ \\ 
        \bottomrule
    \end{tabular}
\end{table}

\clearpage

\newpage

\section*{Supplementary Note 2: Additional surface band structure}

\begin{figure}[ht]
    \centering
    \includegraphics{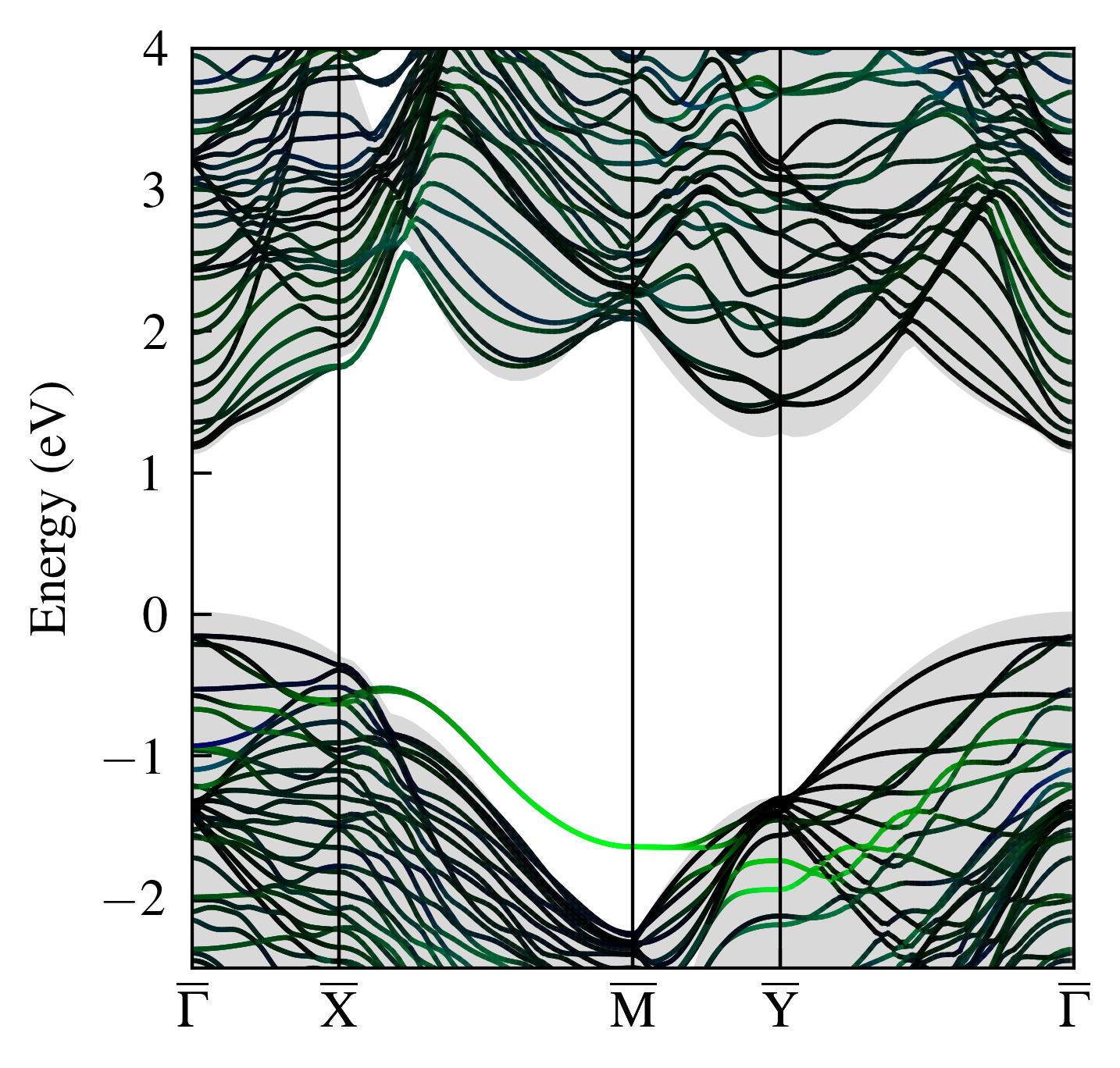}
    \caption{
    Surface band structure of the Si(100)-$(2\times 1)$:As-Si-H surface reconstruction calculated with the hybrid functional HSE06 (see Methods in the main text) as shown in Fig.~2g of the main text, but with different coloring (see text). 
    The $(2\times 1)$ surface Brillouin zone shown in Fig.~2c of the main text.
    }
    \label{fig:additional_bs}
\end{figure}

Supplementary~Fig.~\ref{fig:additional_bs} shows the surface band structure of the Si(100)-$(2\times 1)$:As-Si-H surface reconstruction, already shown in Fig.~2g of the main text, but with a different coloring. 
Here the green color of the bands corresponds to $\sum_{lm}C_{lm, n\mathbf{k}}^{\mathrm{As}}$, i.e., how much the As atom contributes to each state $\vert n\mathbf{k}\rangle$.
The faint blue color visible in some bands corresponds to $\sum_{lm}[C_{lm, n\mathbf{k}}^{\mathrm{Si@S}} + C_{lm, n\mathbf{k}}^{\mathrm{H@S}}]$, i.e., the Si and H atoms directly on the surface (S).
Note that if, for example, the As and Si-H both contribute a state, the colors will mix, i.e., blue and green will become teal.

Looking at Supplementary~Fig.~\ref{fig:additional_bs}, we can observe that the surface state propagating in the bulk band gap along the high-symmetry lines from the $\overline{\mathrm{X}}$ through the $\overline{\mathrm{M}}$ to the $\overline{\mathrm{Y}}$ point originates from the As atom in the Si(100)-$(2\times 1)$:As-Si-H surface reconstruction.

\clearpage

\newpage

\section*{Supplementary Note 3: Changes to the VASP Code}

In this Supplemental Note, we describe the changes made to the source code of VASP version 5.4.4 to tailor the output given after solving the Bethe-Salpeter equation (BSE) of to enable the RAS analysis presented in the main text.
Since our post-processing routines do not replicate the internal symmetry operations of VASP, all calculations had to be performed without symmetry (\texttt{ISYM=-1}). 
This avoids inconsistencies between the BSE eigenvectors and transition matrix elements stored internally by VASP during a calculation and those used by our post-processing routines, i.e., data read from VASP output files, thus ensuring that the exported data remains consistent.

First, we describe our tailored output format.
In version 5.4.4 of VASP, after diagonalizing the BSE Hamiltonian, the eigenvalues $E^\Lambda$ and associated eigenvectors $A^\Lambda_{vc\mathbf{k}}$ for excitons $\Lambda$ up to the index set by \texttt{NBSEEIG} are written to an ASCII file called \texttt{BSEFATBAND}.
When \texttt{NBSEEIG} is greater than 100 or 1000, or when the size of the Hamiltonian is large, the \texttt{BSEFATBAND} file becomes huge and difficult to handle effectively for post-processing purposes using programming languages such as Python.
To avoid this, we have designed the following output format: After the BSE Hamiltonian has been diagonalized, a new directory is automatically created in which individual binary files are stored for all excitons $\Lambda$ which satisfy the condition $\vert O^\Lambda_{xx} - O^\Lambda_{yy} \vert > \mathcal{T}$ (the definition of the oscillator strength $O^\Lambda_{ii}$ is given in the main text).
Here $\mathcal{T}$ is a threshold set to the average contribution of all excitons within the RAS energy range, calculated within the VASP code, specifically in our modified output function, and set to the following values:
$\mathcal{T}_\mathrm{Si(100):P} \sim 6.5\cdot 10^{-2}$, $\mathcal{T}_\mathrm{Si(100):As} \sim 6.5\cdot 10^{-2}$, $\mathcal{T}_\mathrm{Si(100):Sb} \sim 6.7\cdot 10^{-2}$, $\mathcal{T}_\mathrm{Si(100):Bi} \sim 7.3\cdot 10^{-2}$ and  $\mathcal{T}_\mathrm{Si(100):As-Si-H} \sim 6.2\cdot 10^{-2}$.
Each binary file is named by its exciton index $\Lambda$ and contains a data matrix with seven columns: three for the k-point coordinates, one for the valence band index $v$, one for the conduction band index $c$, and two for the real and imaginary parts of the BSE eigenvector $A^\Lambda_{vc\mathbf{k}}$. 
The matrix has $N + 1$ rows, where $N$ is the product of the number of valence bands, number of conduction bands, and k-points included in the calculation (i.e., the length of $A^\Lambda_{vc\mathbf{k}}$).
The first row stores metadata, including the oscillator strength values along the $x$ and $y$ directions and the corresponding BSE eigenvalue.

Second, we describe which parts of the code were adjusted for this modification. 
In the \texttt{mpi.F} file, inspired by a post in the VASP user forum, we changed a 4-byte integer variable to an 8-byte type to avoid integer overflows when dealing with very large BSE matrices, which can easily occur when solving BSE for surfaces.
The details can be read here: \url{https://ww.vasp.at/forum/viewtopic.php?p=21173}.
The \texttt{bse.F} file has undergone more extensive changes, mainly within the \texttt{WRITE\_BSE} subroutine. 
We added a new variable that tracks the anisotropy in oscillator strength for each exciton, allowing us to monitor the difference in oscillator strength in the $x$ and $y$ directions.
The code then loops through all \texttt{NBSEEIG} excitons, and if the anisotropy of, e.g., exciton $\Lambda$ exceeds the chosen threshold $\mathcal{T}$, the code creates the binary file described above for that exciton and stores its information.
Except for these targeted changes, all other aspects of the VASP code were left untouched.

\begin{figure}[ht]
    \centering
    \includegraphics{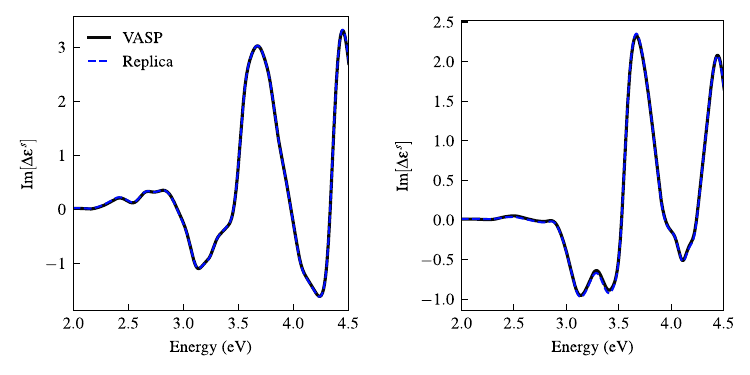}
    \caption{
    Comparison of the imaginary part of the dielectric anisotropy obtained from VASP (solid black line) with the reconstruction using our output files containing only a subspace of optically active excitons (dashed blue line).
    Left panel: Si(100):As. 
    Right panel: Si(100):As-Si-H.
    }
    \label{fig:eps_replica}
\end{figure}

To verify that our modifications accurately reproduce the optical properties derived from the BSE calculations, we performed two different consistency checks.
First, we confirmed that our output format correctly stored the BSE eigenvectors $A^\Lambda_{vc\mathbf{k}}$ by recalculating the value of the stored oscillator strengths $O^\Lambda_{xx}$ and $O^\Lambda_{yy}$ using the BSE eigenvectors stored in the generated binary files in Python. 
Specifically we took the BSE eigenvector $A^\Lambda_{vc\mathbf{k}}$ from the binary file and the transition matrix elements $M^i_{vc\mathbf{k}}=\bra{c\mathbf{k}}\hat{v}_i\ket{v\mathbf{k}}$ from the \texttt{WAVEDER} file to calculate the oscillator strengths $O^\Lambda_{xx}$ and $O^\Lambda_{yy}$, and found that they matched the internally calculated VASP values to numerical accuracy, down to differences on the order of $10^{-12}$.
Secondly, while VASP calculates the dielectric function internally, we repeated this calculation using our output files to ensure that both approaches yield approximately the same spectra, as we only output a subspace of optically active excitons. 
Looking at Supplementary~Fig.~\ref{fig:eps_replica} (same as the upper panel in Fig.~3a and d in the main), the close agreement between the VASP calculation and our reconstruction confirms that extracting a subset of excitons does not compromise the accuracy of the obtained dielectric functions.
The excellent agreement in both tests shows that our custom output routines are working properly.

\clearpage

\newpage

\section*{Supplementary Note 4: Calculations of additional Si(100):V surface reconstructions}

In this Supplementary Note, we present additional analyses of Si(100):V surfaces where the symmetric As dimer Si(100):As surface reconstruction is replaced by dimers of other group V elements (P, Sb, Bi).
We provide an illustration of the structural difference after geometry relaxation, calculated surface-projected density of states, surface band structures, and normalized layer-resolved exciton localization (NLREL) maps for two sets of systems: (i) Si(100):V@As, where the geometry is fixed to that of the Si(100):As reconstruction, and (ii) fully relaxed Si(100):V surfaces.
All calculations were performed using the same methods and calculation parameters as presented in the main text.

\begin{figure}[ht]
    \centering
    \includegraphics{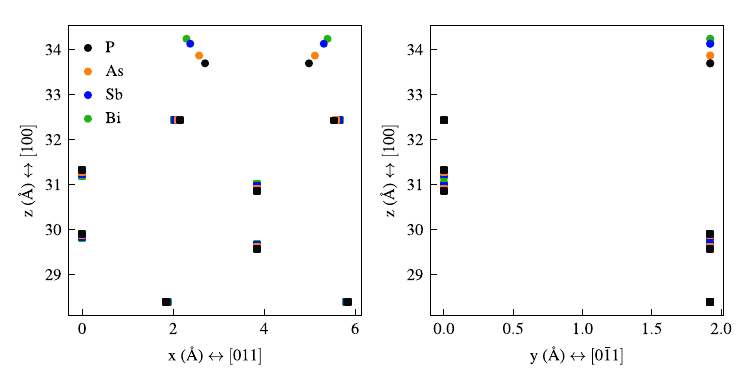}
    \caption{
    Extended version of the structure of the top layers of the relaxed Si(100):V systems shown in Fig.~6 of the main text
    Square markers denote Si atoms, while circular markers denote group V atoms, as indicated by the color coding, which is consistent across all panels.
    Left panel: View of the $xz$ plane spanned by $[011]$ and $[100]$ vectors.
    Right panel: View of the $yz$ plane spanned by $[0\bar{1}1]$ and $[100]$ vectors.
    }
    \label{fig:structure}
\end{figure}

\begin{figure}[ht]
    \centering
    \includegraphics{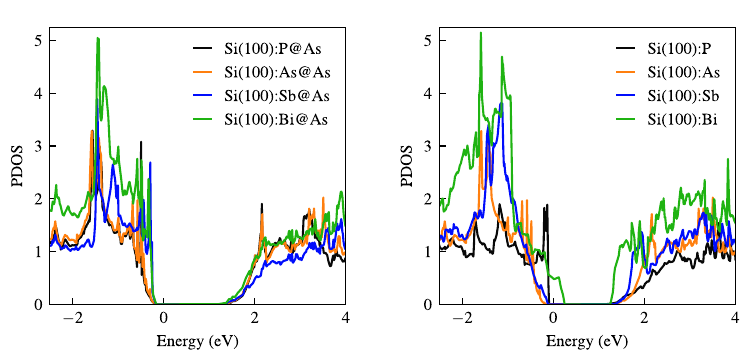}
    \caption{
    Surface-projected density of states calculated using the HSE06 functional.
    Left panel: Projections on the symmetric surface dimers for the Si(100):V systems with fixed Si(100):As geometry.
    Right panel: Projections on the symmetric surface dimers for the relaxed Si(100):V systems.
    }
    \label{fig:dos}
\end{figure}

\begin{figure}[ht]
    \centering
    \includegraphics{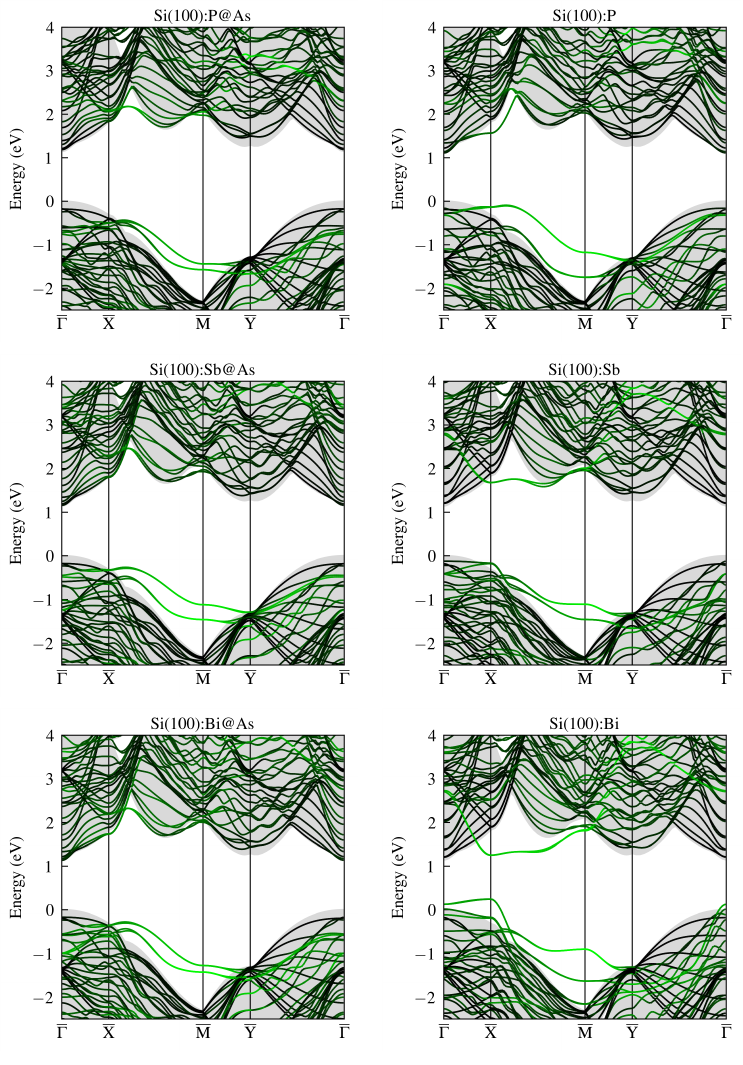}
    \caption{
    Surface band structures calculated using the HSE06 functional for 20-layer slabs. 
    Left column: Si(100):V surface reconstruction with fixed Si(100):As geometry.
    Right column: Relaxed Si(100):V surface reconstruction.
    The $(2\times 1)$ surface Brillouin zone is shown in Fig.~2\textbf{c} in the main text.
    The color of the bands corresponds to  $\sum_{\mathrm{V}}\sum_{lm}C_{lm, n\mathbf{k}}^{\mathrm{V}}$, i.e., the projection on the symmetric surface dimers, where green indicates a high dimer contribution and black indicates no dimer contribution (same as in Fig.~2 of the main text for the Si(100):As reconstruction).
    }
    \label{fig:bandstructures}
\end{figure}

\begin{figure}[ht]
    \centering
    \includegraphics{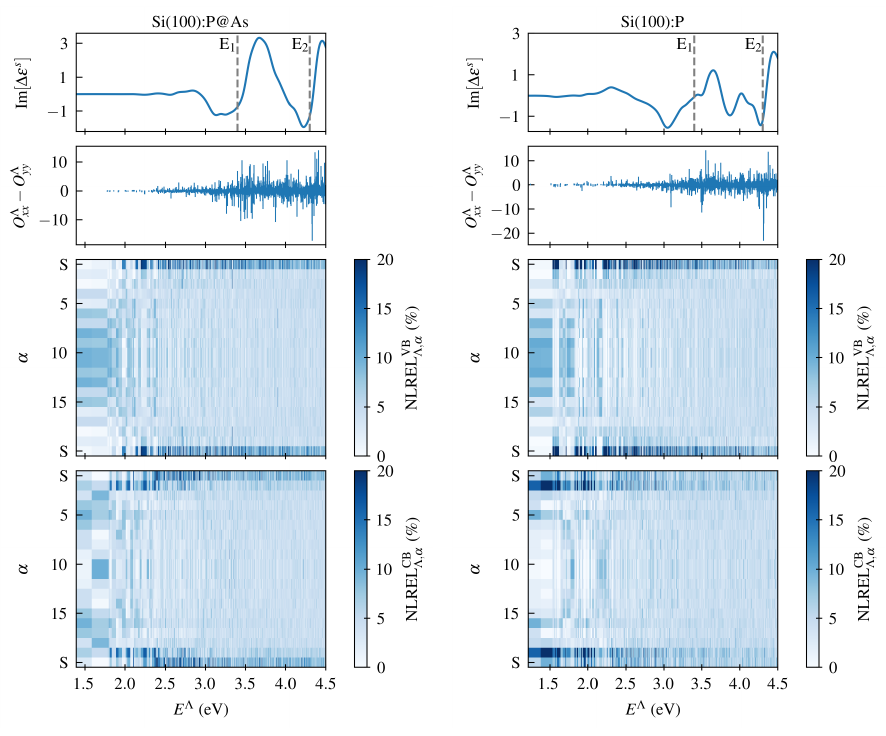}
    \caption{
    Illustration of the surface dielectric anisotropy, oscillator strengths (upper panels), and NLREL maps calculated using Eqs.~(6--7) from the main text (lower panels) for the VB and CB manifolds, respectively.
    Left column: Si(100):P@As surface reconstruction with fixed Si(100):As geometry.
    Right column: Relaxed Si(100):P surface reconstruction.
    In the upper panels, the critical points of the imaginary part of the bulk dielectric function of silicon are marked by vertical dashed gray lines \cite{Lautenschlager1987}.
    We limit the color bars to $20$\% to facilitate comparison between systems.
    }
    \label{fig:excitons_p}
\end{figure}

\begin{figure}[ht]
    \centering
    \includegraphics{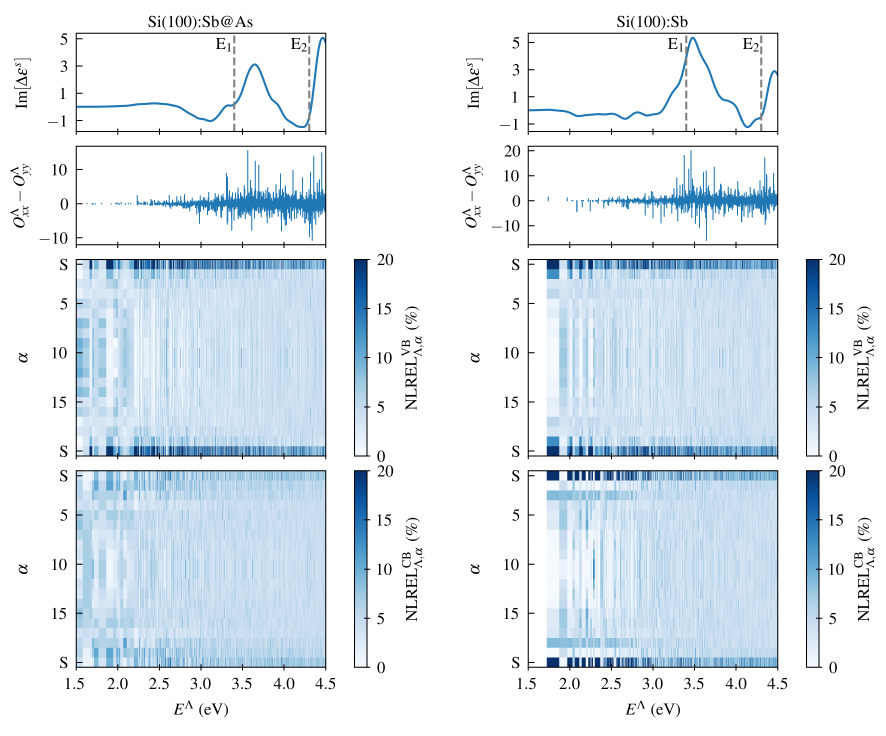}
    \caption{
    Same as Supplementary~Fig.~\ref{fig:excitons_p} but for the constrained and relaxed Si(100):Sb surface reconstructions.
    }
    \label{fig:excitons_sb}
\end{figure}

\begin{figure}[ht]
    \centering
    \includegraphics{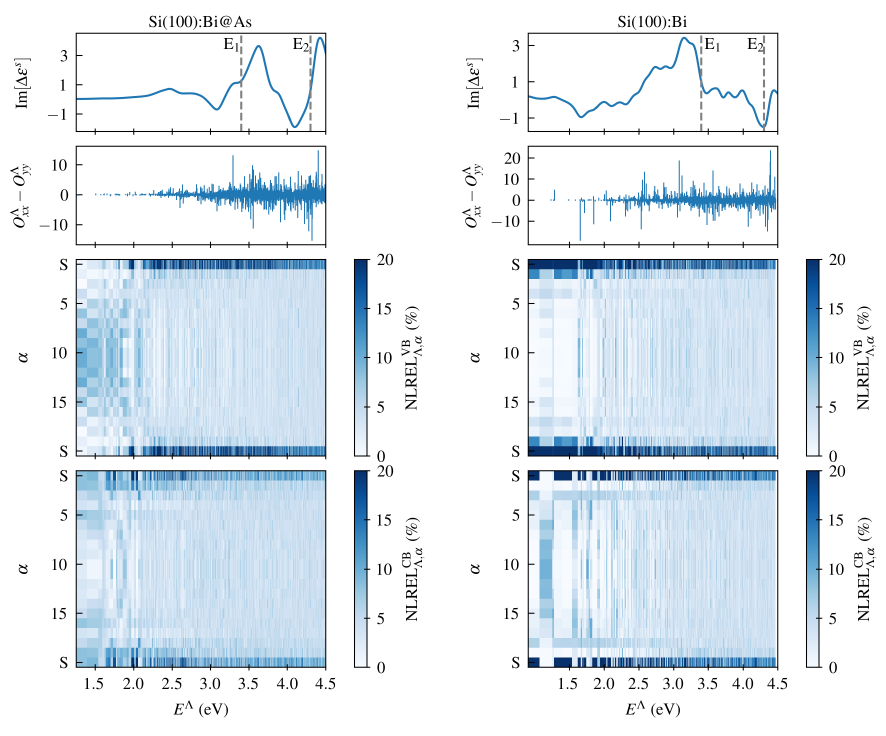}
    \caption{
    Same as Supplementary~Fig.~\ref{fig:excitons_p} but for the constrained and relaxed Si(100):Bi surface reconstructions.
    }
    \label{fig:excitons_bi}
\end{figure}

\clearpage

\newpage

\section*{Supplementary Note 5: Static screening model}

In order to make the calculation computationally tractable and to provide internal consistency between the optical properties calculated for bulk silicon and the various surfaces, we used an analytical model dielectric function when calculating the static screening for the BSE Hamiltonian. 
The model dielectric function has been used successfully in previous studies, e.g., Ref.~\cite{Schmidt2000, Hahn2001, Bokdam2016, Liu2018, Tal2020}, and has the following form \cite{Bechstedt1992}:
\begin{equation}
    \label{eqn:model_screening}
    \varepsilon_{\mathbf{G}\mathbf{G}}^{-1}(\mathbf{q}) = 1 - (1 - \varepsilon_\infty^{-1}) \exp\left({-\frac{\vert \mathbf{q} + \mathbf{G} \vert}{4\mu^2}}\right)
\end{equation}
To obtain $\varepsilon_\infty$, we calculated the macroscopic ion-clamped static dielectric tensor from the response of bulk silicon to finite electric fields (finite difference approach) using the HSE06 functional and obtained $\varepsilon_\infty = 11.115$.
We then calculated $\mu$ by fitting Eq.~(\ref{eqn:model_screening}) to $\varepsilon_{\mathbf{G}\mathbf{G}}^{-1}$ from a G$_0$W$_0$ calculation and obtained $\mu = 1.22$, see Supplementary~Fig.~\ref{fig:model_screening}.
The starting point for the G$_0$W$_0$ calculation were Kohn-Sham energies and wavefunctions from the previous HSE06 calculation. 

\begin{figure}[ht]
    \centering
    \includegraphics{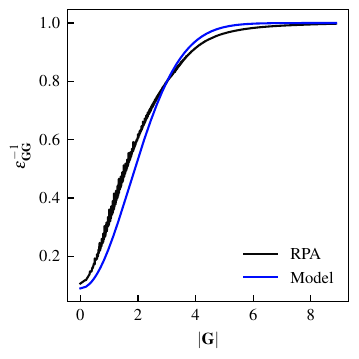}
    \caption{
    Model dielectric function (blue) according to Eq.~(\ref{eqn:model_screening}) fitted to the dielectric function calculated in the random phase approximation (RPA) from a G$_0$W$_0$ calculation (black).
    }
    \label{fig:model_screening}
\end{figure}

\clearpage

\newpage

\section*{Supplementary Note 6: RAS temperature estimation}

In this Supplementary Note, we provide details on the temperature estimation of our low-temperature (LT) RAS measurements of the Si(100)-$(2 \times 1)$-As-Si-H surface reconstruction, prepared by metal organic chemical vapor deposition (MOCVD).

In the ultra-high vacuum (UHV) environment, in particular when transferring the sample from the MOCVD to the cryostat via an MOCVD-to-UHV shuttle \cite{Hannappel2004}, the cryostat's built-in thermal sensor cannot be attached directly to the sample, thus preventing real-time temperature monitoring during LT-RAS measurements.
To address this limitation, we performed a preliminary test in which the thermal sensor was attached to the sample with thermal paste under ambient conditions.
Note that part of the cryostat chamber had to be disassembled to mount the sensor. 
A schematic view of the prepared sample with the thermal sensor attached inside the cryostat is shown in Supplementary~Fig.~\ref{fig:temp_estim}a.
Once the sample was in place, the cryostat was turned on, and after about $2$~hours the copper cooling platform reached a temperature of about $5$~K, see the dashed blue line in Supplementary~Fig.~\ref{fig:temp_estim}b. 
Under these conditions the sample temperature stabilized at about $23$~K, see Supplementary~Fig.~\ref{fig:temp_estim}c. 
To evaluate the influence of the xenon lamp (XBO lamp) used for RAS measurements on the sample temperature, we illuminated a part of the sample not covered by the sensor as shown in Supplementary~Fig.~\ref{fig:temp_estim}a and observed that the lamp caused a negligible temperature variation.

\begin{figure}[ht]
    \centering
    \includegraphics{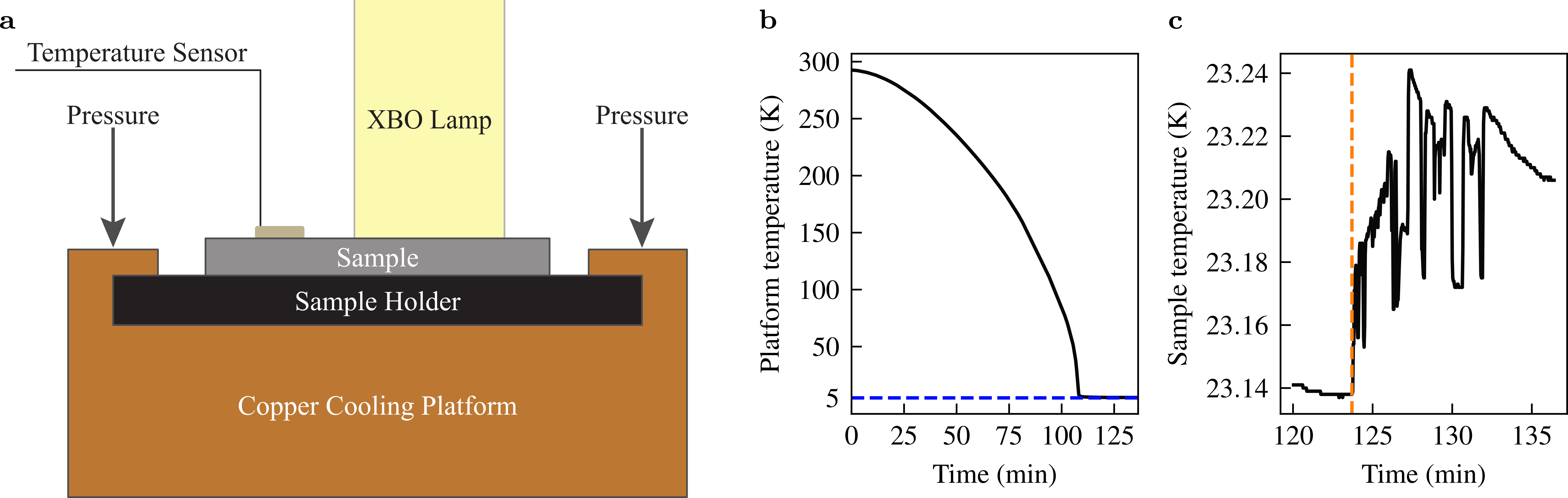}
    \caption{
    Estimation of sample temperature in the cryostat.
    \textbf{a} Schematic of the sample setup: The thermal sensor is attached to the silicon sample with thermal paste, while a portion of the sample remains open and is illuminated by the xenon (XBO) lamp used during RAS measurements to verify that its thermal effect on the sample temperature is negligible.
    The thermal contact between the copper cooling platform and the sample holder is enhanced by the pressure exerted by the clamps, as indicated by the arrows.
    \textbf{b} Temperature evolution of the copper cooling platform after turning on the cryostat, reaching around $5$~K (see dashed blue line) after $2$~hours, starting from ambient conditions. 
    \textbf{c} Sample temperature after running the cryostat for $2$~hours.
    The sample temperature stabilizes at around $23$~K. 
    The dashed orange line indicates the time when the XBO lamp was turned on.
    }
    \label{fig:temp_estim}
\end{figure}

For the actual RAS measurement shown in the main text, the experimental procedure was as follows to minimize surface contamination.
Prior to sample transfer, the cryostat chamber was pumped to UHV and precooled overnight.
This ensured a pressure inside the cryostat of less than $5\cdot 10^{-9}$~mbar and a stable platform temperature of $5$~K prior to sample insertion. 
The next day, the sample was transferred into the chamber via an MOCVD-to-UHV shuttle \cite{Hannappel2004}.
Before measuring the RAS, we waited $15$~minutes to allow the sample temperature to stabilize (cf.~Ref.~\cite{Visbeck2001}).
However, due to the challenges associated with the transfer, the thermal contact between the copper cooling block and the sample holder was likely inferior to that in the preliminary test at ambient conditions, resulting in an estimated sample temperature in the vicinity of $50$~K.

Since direct temperature estimation was not possible during LT RAS measurements in the cryostat, we also attempted to estimate the sample temperature in real time using an optical method.
Visbeck~\textit{et~al.}~\cite{Visbeck2001} have analyzed the temperature dependence of the RAS for several InP(100) surface reconstructions, providing a reference dataset for temperature calibration. 
There, the energy dependence of the characteristic RAS features as a function of temperature is described by
\begin{align}
    E(T) = E_0 - \alpha \left(1 + \frac{2}{e^{(\theta/T)}-1}\right)
\end{align}
where $E(T)$ is the temperature-dependent peak position, and the reconstruction and peak-dependent fit parameters $E_0$, $\alpha$, and $\theta$ can be found in Tab.~1 of Ref.~\cite{Visbeck2001}. 
To extract the temperature from a measured peak energy, the equation is rewritten as
\begin{align}
    \label{eqn:ras_temp}
    T = \theta \cdot \left[\ln\left(1+\frac{2}{\frac{E_{0}-E(T)}{\alpha}-1}\right)\right]^{-1}.
\end{align}
We used an InP(100) sample with a P-rich ($2 \times 1$) surface reconstruction, prepared by MOCVD and subsequently transferred to the cryostat environment in UHV, again via an MOCVD-to-UHV shuttle \cite{Hannappel2004}. 
To investigate possible changes in the RA spectra with longer cooling times in the cryostat, we measured the RAS evolution of our sample over an extended waiting period in the cryostat.
The results are shown in Supplementary~Fig.~\ref{fig:InP_temp_evo}. 
Over the course of one day, we observed a shift of the $A$ peak from about $1.97$~eV to $2.11$~eV, which we attribute to the adsorption of residual water molecules on the surface. 
This effect is consistent with previous studies investigating the influence of water on the RAS of InP(100) and GaInP(100) surfaces during controlled exposure experiments \cite{May2017expo, Ostheimer2024}.

\begin{figure}[ht]
    \centering
    \includegraphics{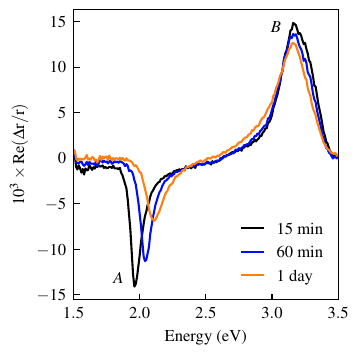}
    \caption{
    Time evolution of the RA spectra of the P-rich InP(100)-($2 \times 1$) surface reconstruction in the cryostat.
    The $A$ peak is blue-shifted over time due to water adsorption, while the $B$ peak remains stable and only slightly decreases in intensity.
    }
    \label{fig:InP_temp_evo}
\end{figure}

In contrast, the $B$ peak at $3.17$~eV remains relatively stable, which is also consistent with previously mentioned studies \cite{May2017expo, Ostheimer2024}, making it a more reliable reference for temperature estimation.
Since surface adsorption effects can affect peak positions, we used the $B$ peak $15$~minutes after insertion of the sample into the cryostat (black line in Supplementary~Fig.~\ref{fig:InP_temp_evo}) to ensure that the sample temperature had stabilized while also ensuring that the peak intensity and position were only slightly affected by water adsorption.
Using Eq.~(\ref{eqn:ras_temp}) and $E(T)=3.17$~eV we obtain a temperature of $150$~K with a large uncertainty of about $50$~K, of which about $30$~K is due to the uncertainty of the fit parameters from Ref.~\cite{Visbeck2001}.

The high and uncertain temperature estimate from the optical method is likely due to the weak temperature dependence of RAS peak positions below $150$~K (cf.~Fig.4 in Ref.\cite{Visbeck2001}), uncertainties in the fit parameters and the high sensitivity of the method to the exact peak position.
For example, shifting the $B$ peak from $3.17$~eV to $3.18$~eV lowers the estimated temperature from $150$~K to $125$~K, a change that could easily result from water adsorption effects during the brief cryostat exposure. 
All things considered, we believe that our direct sensor-based estimate of about $50$~K is more reliable, even though it was not obtained simultaneously with the RAS measurement.

\begin{figure}[ht]
    \centering
    \includegraphics{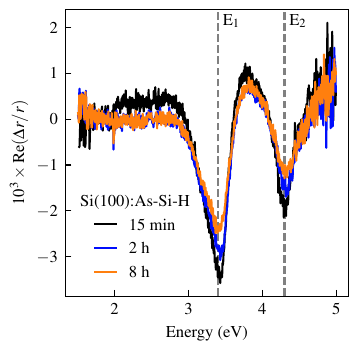}
    \caption{
    Time evolution of the RA spectra of the Si(100)-$(2 \times 1)$-As-Si-H surface reconstruction in the cryostat.
    }
    \label{fig:ras_colorplot}
\end{figure}

As seen in the case of InP(100) in Supplementary~Fig.~\ref{fig:InP_temp_evo}, the likely adsorption of water affects the surface reconstruction and thus the measured RAS the longer the sample remains in the cryostat.
Therefore, in Supplementary~Fig.~\ref{fig:ras_colorplot} we compare RA spectra of Si(100):As-Si-H recorded for longer cooling times inside the cryostat, similar to Supplementary~Fig.~\ref{fig:InP_temp_evo}). 
In contrast to InP(100), the peak positions remain stable over time and only a slight decrease in intensity is observed, indicating a higher resistance to water adsorption compared to InP(100).
We note that this resistance may be attributed to the findings discussed in the main text, i.e., that the states contributing to both features in the RA spectrum are not localized directly at the surface, causing the effect of water adsorption to be small.
Similar to the InP(100) temperature estimate, the spectrum shown in the main text corresponds to the black curve recorded $15$~minutes after insertion to ensure that the sample temperature had stabilized while also ensuring that the peak intensity was only slightly affected by water adsorption.

\begin{figure}[ht]
    \centering
    \includegraphics{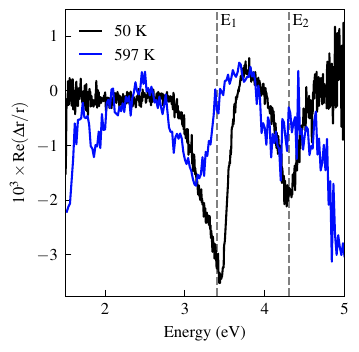}
    \caption{
    Comparison of Si(100):As-Si-H spectra recorded at around $50$~K (cryostat sample temperature) and at $597$~K (MOCVD process sample temperature).
    }
    \label{fig:ras_hightemp}
\end{figure}

Comparing the RA spectrum of the Si(100)-$(2 \times 1)$-As-Si-H surface reconstruction at about $50$~K (cryostat sample temperature) and at $597$~K (MOCVD process sample temperature) in Supplementary~Fig.~\ref{fig:ras_hightemp}, we clearly observe sharper and less noisy blue-shifted features at low temperatures, highlighting the importance of LT measurements.
The reasons for the peak renormalization have already been discussed in detail in the main text.

\clearpage

\newpage

\section*{Supplementary Note 7: Assessment of sample contamination after low temperature measurement}

To ensure that the aforementioned water contamination in the cryostat did not induce structural or chemical changes in the sample, we performed additional low-energy electron diffraction (LEED) and X-ray photoelectron spectroscopy (XPS) measurements after the LT RAS measurements. 

\begin{figure}[ht]
    \centering
    \includegraphics{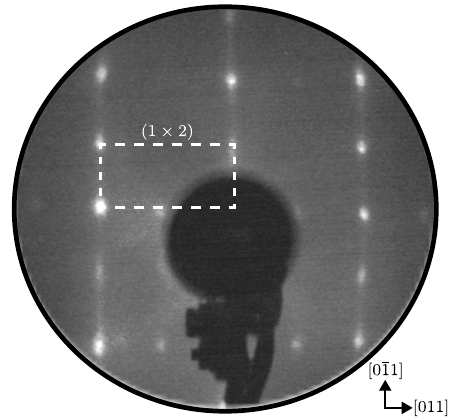}
    \caption{
    LEED pattern after LT RAS measurements confirming a $(1 \times 2)$ majority domain with a weak signal for the $(2 \times 1)$ minority domain.
    }
    \label{fig:leed}
\end{figure}

Before discussing the LEED measurement, we want to quickly emphasize that the difference between $(1\times 2)$ and $(2\times 1)$ affects the orientation of the dimers either along or perpendicular to the step edges of a vicinal surface, cf.~Fig.~1 of Ref.~\cite{Paszuk2018b}.
Throughout the rest of the Supplementary Information and the main text, we have always referred to a $(2\times 1)$ reconstruction, since all \textit{ab initio} calculations have been performed for non-vicinal surfaces, making this distinction irrelevant until now, i.e., when discussing the LEED.
The LEED pattern in Supplementary~Fig.~\ref{fig:leed} confirms that our A-type Si(100):As-Si-H sample retains its characteristic $(1 \times 2)$ reconstruction after LT RAS measurements.
A clear half-order diffraction spot along the $[0\overline{1}1]$ direction indicates a $(1 \times 2)$ surface reconstruction with dimer rows predominantly aligned parallel to the step edges. 
A weak $(2 \times 1)$ minority domain, with dimer rows oriented perpendicular to the step edges, is also visible, but its intensity cannot be quantitatively determined from LEED alone, see Ref.~\cite{Bohlemann2024}.
The agreement between the LEED and RAS results suggests that no significant structural changes or contamination-induced disorder occurred during the cryogenic measurements, i.e., the surface reconstruction remained stable.

\begin{figure}[ht]
    \centering
    \includegraphics{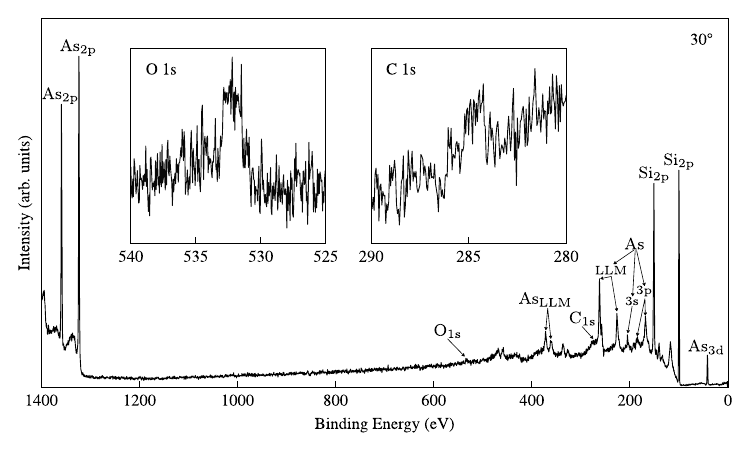}
    \caption{
    XPS survey spectrum of the Si(100)-$(2 \times 1)$-As-Si-H surface after LT RAS with insets for the O~1s and C~1s peaks.
    }
    \label{fig:xps_survey}
\end{figure}

\begin{figure}[ht]
    \centering
    \includegraphics{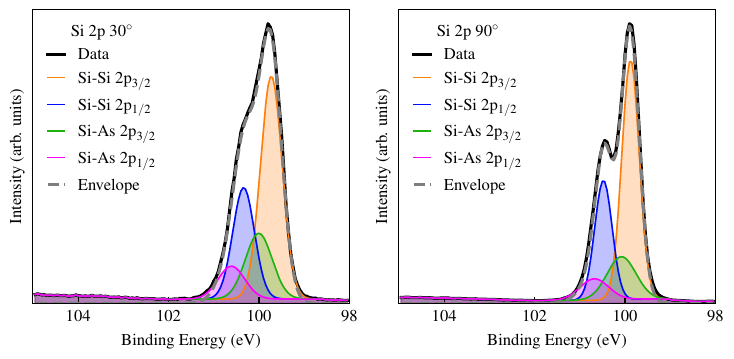}
    \caption{
    XPS scans of the Si~2p core levels.
    Left panel: Measured at a $30^\circ$ photoelectron takeoff angle.
    Right panel: Measured at a $90^\circ$ photoelectron takeoff angle.
    Both measurements (black) were fitted with components related to Si-Si (blue and orange) and Si-As (green and magenta) bonds.
    The components related to Si-As bonds are increased for the $30^{\circ}$ measurement compared to those for the $90^{\circ}$ measurement.
    }
    \label{fig:xps_si2p}
\end{figure}

To further investigate possible contamination, we measured an XPS survey spectrum with a $30^{\circ}$ takeoff angle for high surface sensitivity, see Supplementary~Fig.~\ref{fig:xps_survey}.
There, one can observe only major peaks corresponding to silicon and arsenic, with negligible contributions from oxygen and carbon (see insets). 
This indicates that the MOCVD surface preparation process effectively removed unwanted precursors and that the UHV transfer was successful in preserving the surface composition, and that the contaminants in the cryostat did not alter the chemical composition of the sample surface.

The Si~2p core level spectrum, analyzed in more detail in Supplementary~Fig.~\ref{fig:xps_si2p}, confirms the absence of significant oxidation. 
The absence of a SiO$_{2}$ peak at $104$~eV indicates that the surface remained free of oxygen-bonded silicon. 
Any minor oxygen traces visible in the insets of the XPS survey in  Supplementary~Fig.~\ref{fig:xps_survey} are likely due to transfer-related exposure rather than intrinsic contamination present in the cryostat.
Measurements taken at $30^{\circ}$ show enhanced contributions from Si-As bonding compared to those taken at $90^{\circ}$ (the intensity ratio changes from $0.038$ to $0.019$), confirming that the topmost atomic layers retain the expected chemical composition.
The XPS spectra shown in Supplementary~Fig.~\ref{fig:xps_si2p} were fitted with Gaussian-Lorentzian functions, keeping the full width at half maximum constant in each group (Si-Si or Si-As).
The intensity ratios of Si~2p$_{3/2}$ to Si~2p$_{1/2}$ were fixed at a 2:1 ratio, with a spin-orbit splitting of $0.61$~eV, in agreement with previous studies \cite{Paszuk2018a, Paszuk2018b, Bohlemann2024}.

\newpage

\clearpage


\providecommand{\noopsort}[1]{}\providecommand{\singleletter}[1]{#1}%